\documentclass[aps,prl,reprint,superscriptaddress,nobalancelastpage]{revtex4-2}

\usepackage{braket}
\usepackage{amsmath}
\usepackage{graphicx}
\usepackage{bm}
\usepackage[colorlinks=true,linkcolor=blue,citecolor=blue,urlcolor=blue]{hyperref}
\newcommand{\tcell}[1]{{\raggedright #1}}

\begin{document}

% Use the \preprint command to place your local institutional report
% number in the upper righthand corner of the title page in preprint mode.
% Multiple \preprint commands are allowed.
% Use the 'preprintnumbers' class option to override journal defaults
% to display numbers if necessary
%\preprint{}

%Title of paper
\title{Analytic Theory of Phase Transitions in Optical Metamaterials}

\author{Jiahao Yan}
% \email{jiahao.yan@wustl.edu}
\affiliation{Department of Physics, Washington University in St. Louis, St. Louis, MO 63130, USA}
\author{Xiaoshui Lin}
\affiliation{Department of Physics, Washington University in St. Louis, St. Louis, MO 63130, USA}
\author{Junpeng Hou}
% \email{jhou@pinterest.com}
\affiliation{Pinterest Inc., San Francisco, California 94103, USA}

\author{Qing Gu}
\affiliation{Department of Electrical and Computer
Engineering, North Carolina State University, Raleigh, North
Carolina 27695, USA}
\affiliation{Department of Physics and Astronomy,
North Carolina State University, Raleigh, North Carolina
27695, USA}

\author{Chuanwei Zhang}
\email{chuanwei.zhang@wustl.edu}
\affiliation{Department of Physics, Washington University in St. Louis, St. Louis, MO 63130, USA}
\affiliation{Center for Quantum Leap, Washington University in St. Louis, St. Louis, MO 63130, USA}

\begin{abstract}
{
Optical metamaterials provide a versatile platform for engineering homogeneous electromagnetic media whose distinct phases are characterized by phase diagrams in constitutive-parameter space. However, existing studies of hyperbolicity, topological properties, and exceptional-point formation often rely on highly symmetric models or case-by-case numerical parameter scans, leaving a unified analytic framework that identifies phases and phase transitions directly from the constitutive tensors lacking. Here, we develop a general theory that yields exact analytic criteria for topological transitions, exceptional-point transitions, pinch-off Lifshitz transitions, and optical Lifshitz transitions in homogeneous media. Applying this framework to a tractable example of a gyroelectric medium with anisotropic chirality, we uncover exceptional rings and negative refraction induced by gyroelectric–chiral coupling. By enabling the exact determination of phase boundaries, our theory provides a predictive framework for discovering previously unexplored electromagnetic phases and offers new principles for the systematic design of optical metamaterials.
}
\end{abstract}

% insert suggested keywords - APS authors don't need to do this
%\keywords{}

%\maketitle must follow title, authors, abstract, and keywords
\maketitle

% body of paper here - Use proper section commands
% References should be done using the \cite, \ref, and \label commands

\textit{Introduction} The advent of optical metamaterials has fundamentally transformed the paradigm of electromagnetic design, enabling the realization of effective media with unprecedented constitutive properties \cite{Veselago:1968, smithCompositeMediumSimultaneously2000,pendryNegativeRefractionMakes2000,shelbyExperimentalVerificationNegative2001,smithElectromagneticWavePropagation2003,poddubnyHyperbolicMetamaterials2013}. Through tailored permittivity, permeability, and chirality tensors, these materials provide a versatile platform for realizing and exploring exotic electromagnetic phenomena governed by the geometry and topology of their dispersion relations in momentum space. In recent years, a variety of distinct phases and associated transitions have been identified, each playing a central role in determining the functionality of these engineered media.

\textbf{Hyperbolic Phases and Optical Lifshitz Transitions.} Hyperbolic metamaterials have attracted considerable attention because they support open equal-frequency surfaces (EFSs) with, in principle, unbounded wave vectors. This distinctive dispersion enables applications ranging from subwavelength imaging to sensing and lasing \cite{poddubnyHyperbolicMetamaterials2013,jacobOpticalHyperlensFarfield2006,liuFarFieldOpticalHyperlens2007,vasilantonakisRefractiveIndexSensing2015,galfskyActiveHyperbolicMetamaterials2015a,jacobBroadbandPurcellEffect2012}. The transition between elliptical and hyperbolic phases is typically described as an optical Lifshitz transition. This transition also alters the photonic density of states, providing a direct route to Purcell-enhanced spontaneous emission \cite{smolyaninovMetricSignatureTransitions2010,reyes-gomezMetricsignatureTopologicalTransitions2014,poddubnyMicroscopicModelPurcell2012,galfskyActiveHyperbolicMetamaterials2015a,jacobBroadbandPurcellEffect2012}.

\textbf{Topological Phases and Topological Phase Transitions.} Topological photonics has enabled robust, defect-immune surface transport \cite{haldanePossibleRealizationDirectional2008,raghuAnalogsQuantumHalleffectEdge2008,wangObservationUnidirectionalBackscatteringimmune2009,khanikaevPhotonicTopologicalInsulators2013,luTopologicalPhotonics2014,slobozhanyukExperimentalDemonstrationTopological2016,yangDirectObservationTopological2017,ozawaTopologicalPhotonics2019,xuMultipleOneWayEdge2019}. By incorporating chirality and gyrotropy, homogenizable metamaterials can host topological phases characterized by non-zero Chern numbers and distinctive surface-state arcs \cite{calozElectromagneticChirality2019,gaoTopologicalPhotonicPhase2015,houTopologicalBandsTriply2020,houTopologicalHyperbolicDielectric2021,chernChiralSurfaceWaves2017,shiuPhotonicChernInsulators2020,liTopologicalHyperbolicMetamaterials2024}.

\textbf{Exceptional Points and EP Transitions.} The exploration of non-Hermitian media has established exceptional points (EPs) as a platform for unconventional wave control. These spectral singularities unlock phenomena such as enhanced parametric sensitivity, bulk Fermi arcs, and unidirectional reflectionless dynamics \cite{el-ganainyNonHermitianPhysicsPT2018,miriExceptionalPointsOptics2019,ozdemirParityTimeSymmetry2019,kaminskiControlExceptionalPoints2017,parkObservationExceptionalPoint2020,zhenSpawningRingsExceptional2015,zhouObservationBulkFermi2018,yangCreatingPairsExceptional2024,ashidaNonHermitianPhysics2020,fengExperimentalDemonstrationUnidirectional2013,huangUnidirectionalReflectionlessLight2017}.

Despite the profound implications of these phases, their current understanding remains largely confined to model-specific configurations with high symmetry. Distinct phases are often identified on a case-by-case basis through extensive parameter scans \cite{parkObservationExceptionalPoint2020,zhenSpawningRingsExceptional2015,gaoTopologicalPhotonicPhase2015,houTopologicalBandsTriply2020,houTopologicalHyperbolicDielectric2021,krishnamoorthyTopologicalTransitionsMetamaterials2012a,shchelokovaMagneticTopologicalTransition2014}. A central unresolved challenge is the absence of a unified, top-down theoretical framework that can predict the phases and their transitions directly from an arbitrary constitutive tensor. Without such a formulation, determining the exact parameter boundaries governing hyperbolic EFSs, topological bands, and the emergence of EPs remains difficult.

In this Letter, we develop a general theory that addresses this challenge by establishing a comprehensive classification of phase boundaries in homogeneous media. Starting from the generalized Maxwell eigenproblem \cite{raghuAnalogsQuantumHalleffectEdge2008,silveirinhaChernInvariantsContinuous2015,silveirinhaBulkedgeCorrespondenceTopological2016}, we derive exact analytical criteria for topological transitions, EP transitions, pinch-off Lifshitz transitions, and optical Lifshitz transitions. For non-dispersive media, the theory admits a direct geometric interpretation in momentum space: intersections of EFSs signal topological phase transitions, whereas rays from the origin tangent to an EFS identify exceptional points or rings. We further show that optical Lifshitz transitions are associated with the emergence of zero-frequency EPs and determine the conditions under which energy-metric singularities coincide with these phase boundaries. To demonstrate the utility of the theory, we analyze a gyroelectric medium with anisotropic chirality and uncover finite-frequency exceptional rings and negative refraction induced by gyroelectric–chiral coupling. By systematically delineating the corresponding phase boundaries, our framework provides a practical guide for designing homogenizable metamaterials.

\textit{Model:}
Within the effective medium approximation for metamaterials, the source-free Maxwell's equations can be cast as a generalized eigenvalue problem \cite{raghuAnalogsQuantumHalleffectEdge2008}: 
\begin{eqnarray}
     H_{P}\psi	&=&\omega H_{M}\psi  \\
\psi=\left(\begin{array}{c}
\mathbf{E}\\
\mathbf{H}
\end{array}\right),\;H_{P}&=&\left(\begin{array}{cc}
0 & -\mathbf{k}\times\\
\mathbf{k}\times & 0
\end{array}\right),	\;H_{M}=\left(\begin{array}{cc}
\epsilon & i\gamma\\
-i\gamma^{\mathsf{T}} & \mu
\end{array}\right)\nonumber
\end{eqnarray}
where $(\mathbf{k}\times)_{ij}:=-\varepsilon_{ijl}k_l$ and $\epsilon$, $\mu$, and $\gamma$ are the permittivity, permeability, and chirality tensors, respectively. 
When $H_M$ is positive-definite, the effective Hamiltonian $H\left(\mathbf{k}\right)=H_{M}^{-1}H_{P}$ is quasi-Hermitian \cite{ashidaNonHermitianPhysics2020}, ensuring a completely real spectrum characteristic of elliptic metamaterials. Non-Hermitian phenomena, including the emergence of EPs, require $H_M$ to become indefinite. By eliminating the magnetic field {$\mathbf{H}$}, we obtain the master equation for the electric field $\mathbf{E}$:
\begin{eqnarray}
    &\bigg[\left(\mathbf{k}\times\right)\mu^{-1}\left(\mathbf{k}\times\right)+i\omega\left((\mathbf{k}\times)\mu^{-1}\gamma^{\mathsf{T}}+\gamma\mu^{-1}(\mathbf{k}\times)\right)\nonumber \\
    &+\omega^{2}\left(\epsilon-\gamma\mu^{-1}\gamma^{\mathsf{T}}\right)\bigg]\mathbf{E}:=K\mathbf{E}=0
    \label{eq:master_eq}
\end{eqnarray}
Nontrivial solutions to the master equation are governed by the dispersion equation $g\left(\omega,\mathbf{k}\right)\equiv\det\left(K\right)=0$. Away from resonant poles, material dispersion becomes negligible, allowing the medium to be effectively treated as non-dispersive, and $g$ reduces to a homogeneous function. After factoring out two static modes, the propagating modes are characterized by $f\left(\omega,\mathbf{k}\right)=g\left(\omega,\mathbf{k}\right)/\omega^2=0$.
At a fixed frequency $\omega$, the set of allowed wavevectors $\mathbf{k}$ satisfying this equation forms the EFS.

\textit{EP Transition, Topological Transition, and Pinch-off Lifshitz transition:} 
A topological phase transition necessarily involves a band crossing, at which distinct eigenmodes become degenerate and the topological band gap closes. By contrast, an exceptional-point (EP) transition is a non-Hermitian phase transition that occurs when system parameters are tuned to or across an EP, causing two or more eigenvalues and their corresponding eigenstates to coalesce defectively. A pinch-off Lifshitz transition, meanwhile, occurs when an entire EFS contracts to a single point in momentum space and subsequently disappears.

These phenomena impose distinct algebraic constraints on the dispersion equation $f(\omega,\mathbf{k})=0$. Originating from geometric singularities in momentum space where the local group velocity becomes ill-defined, they dictate the failure of the implicit function theorem, requiring $f=\partial_\omega f=0$ at a finite frequency $\omega \neq 0$. The corresponding transitions are then classified according to the spatial gradient $\nabla_{\mathbf{k}}f$. When $\nabla_{\mathbf{k}}f=\mathbf{0}$, singularity is a stationary point, where the local EFS geometry is determined by the Hessian matrix $f_{\mathbf{k}\mathbf{k}}$. An indefinite $f_{\mathbf{k}\mathbf{k}}$ produces a conical intersection, implying a band crossing that is necessary for a topological transition, while a definite $f_{\mathbf{k}\mathbf{k}}$ gives rise to a pinch-off Lifshitz transition. By contrast, an EP corresponds to defective mode coalescence and can emerge when the spatial gradient is nonzero ($\nabla_{\mathbf{k}}f \neq \mathbf{0}$). Detailed derivations for these criteria are provided in the Supplemental Material \cite{SM}.
\begin{figure}[t]
    \centering
    \includegraphics[width=1\linewidth]{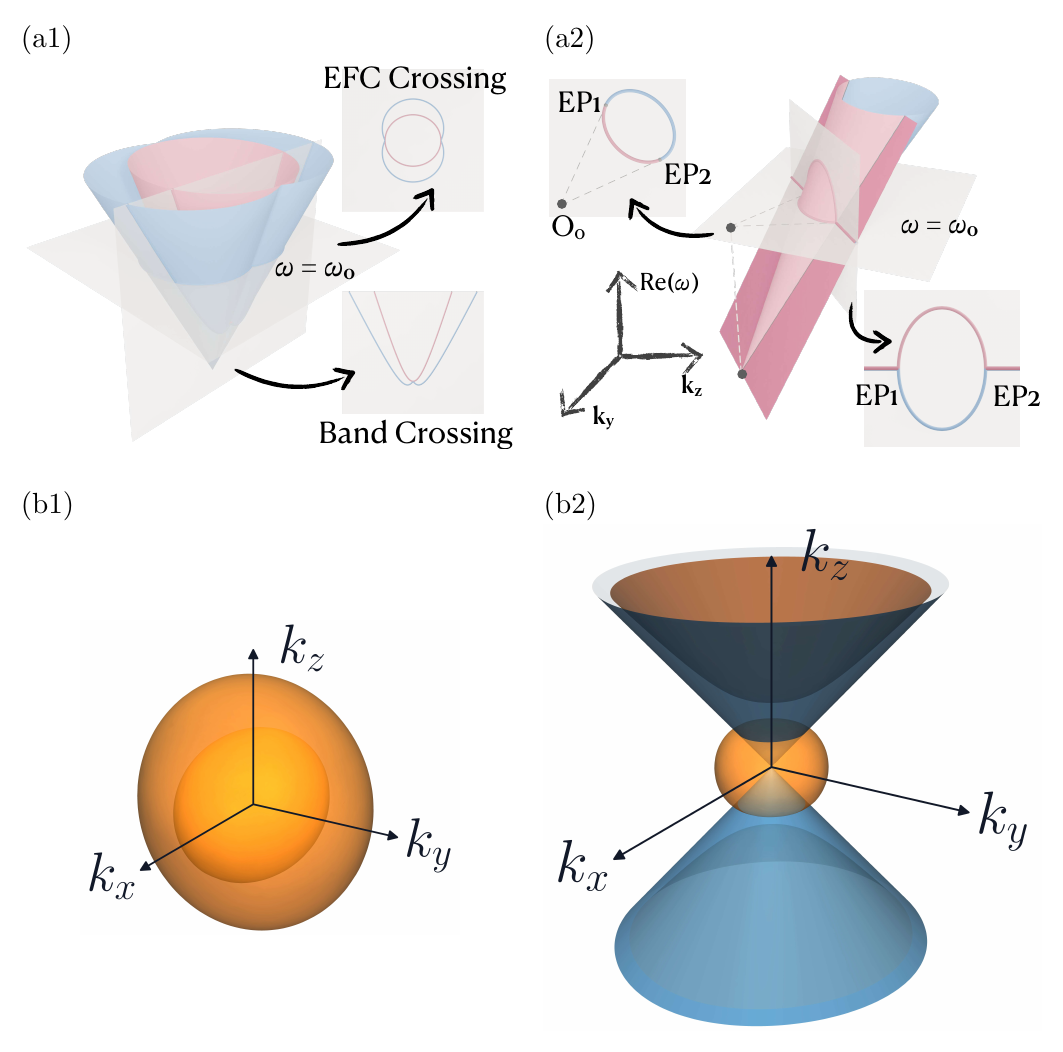}
    \caption{(a1, a2) Geometric signatures of band degeneracies in a 2D non-dispersive medium, with the vertical axis denoting the real part of the frequency, $\text{Re}(\omega)$. The light pink and blue surfaces represent energy bands with purely real frequencies. (a1) A band crossing manifests as a conical intersection, projecting onto the $\omega = \omega_0$ plane as intersecting EFCs. (a2) An EP emerges at a tangency point between the EFC and a ray from the origin ($\mathrm{O_0}$). At these tangency points ($\mathrm{EP_1}$, $\mathrm{EP_2}$), the eigenmodes coalesce. Beyond the EPs, the spectrum becomes complex: the two branches become degenerate in their real parts (visualized as a single, overlapping dark-pink surface), while bifurcating in their imaginary parts. (b1, b2) Illustration of the EFSs across the optical Lifshitz transition. (b1) Before the transition ($\epsilon=\mathrm{diag}(4,4,3)$, $\mu=I_{3\times3}$, $\gamma=\mathrm{diag}(1,0,0)$), all EFSs are closed ellipsoids (visualized as the yellow surfaces). (b2) After the transition at $\epsilon_z=0$ ($\epsilon=\mathrm{diag}(4,4,-3)$, $\mu=I_{3\times3}$, $\gamma=\mathrm{diag}(1,0,0)$), one EFS branch transitions into an open hyperboloid. The blue asymptotic cone corresponds to the zero frequency EP cone governed by $F(\mathbf{k})=0$.}
    \label{fig:DP_EP}
\end{figure}

\begin{table*}[t]
  \caption{\label{tab:transitions}
  Summary of different transitions}
  \begin{ruledtabular}
  \renewcommand{\arraystretch}{1.15} 
  \begin{tabular}{p{0.22\textwidth} p{0.31\textwidth} p{0.45\textwidth}}
    \textbf{Transition Type} & \textbf{Phenomenon} & \textbf{Critical Condition(s)} \\
    \colrule
    
    \tcell{Topological transition} &
    \tcell{EFS crossing (Chern number changes)} &
    \tcell{$\omega \neq 0:\ f = \partial_\omega f = \partial_{k_i} f = 0$, $f_{\mathbf{k}\mathbf{k}}$ indefinite  (Necessary)} \\
    
    \tcell{Pinch-off Lifshitz transition} &
    \tcell{EFS collapse (disappears $\leftrightarrow$ appears)} &
    \tcell{$\omega \neq 0:\ f = \partial_\omega f = \partial_{k_i} f = 0$, $f_{\mathbf{k}\mathbf{k}}$ definite} \\
    
    \tcell{EP transition} &
    \tcell{Mode coalescence (EPs annihilate \newline $\leftrightarrow$ emerge)} &
    \tcell{$\omega \neq 0:\ f = \partial_\omega f = 0$, $\nabla_{\mathbf{k}} f \neq \mathbf{0}$ (Sufficient)\newline
    ($\implies \mathbf{k}\cdot\nabla_{\mathbf{k}}f=0$ for non-dispersive)\newline
    $\omega=0:\ \min_{\hat{\mathbf{k}}} F(\hat{\mathbf{k}}) = 0 \quad \text{or} \quad \max_{\hat{\mathbf{k}}} F(\hat{\mathbf{k}}) = 0$} \\
    
    \tcell{Optical Lifshitz transition} &
    \tcell{EFS geometry change (Closed $\leftrightarrow$ Open)} &
    \tcell{$\min_{\hat{\mathbf{k}}} F(\hat{\mathbf{k}}) = 0 \quad \text{or} \quad \max_{\hat{\mathbf{k}}} F(\hat{\mathbf{k}}) = 0$} \\
    
    \tcell{Energy-metric singularity} &
    \tcell{Quasi-Hermitian $\leftrightarrow$ Non-Hermitian} &
    \tcell{$\det(H_{M})=0$} \\
  \end{tabular}
  \end{ruledtabular}
\end{table*}

As shown in Fig.~\ref{fig:DP_EP}, the critical conditions for topological  and EP transitions admit distinct  geometric interpretations in terms of equal-frequency contours (EFCs). The band crossing condition is equivalent to an EFS intersection [Fig.~\ref{fig:DP_EP}(a1)]. For non-dispersive media, the homogeneity of $f$ implies, by Euler's theorem, $4f=\omega\partial_{\omega}f+\mathbf{k}\cdot\nabla_{\mathbf{k}}f$. At an EP, this relation converts the constraint $\partial_{\omega}f=0$ into $\mathbf{k}\cdot\nabla_{\mathbf{k}}f=0$, requiring the wavevector $\mathbf{k}$ to be orthogonal to the EFC normal $\nabla_{\mathbf{k}}f$. Therefore, EPs occur precisely where the EFC is tangent to a ray emanating from the origin ($O_0$) [Fig.~\ref{fig:DP_EP}(a2)]. Beyond these tangency points, the real parts of the eigenfrequencies merge into a degenerate surface, whereas their imaginary parts bifurcate.

\textit{EP at $\omega=0$ and the Optical Lifshitz Transition:}
At $\omega=0$, the generalized eigenproblem reduces to $H_P \psi = 0$, supporting two longitudinal static modes. A zero-frequency EP arises when an additional mode coalesces with these static states. Assuming a nonsingular permeability tensor $\mu$, we extract a characteristic polynomial $F(\mathbf{k})$ from the leading-order of the dispersion equation $f(\omega,\mathbf{k})$ in the large-$k$ limit \cite{SM}:
\begin{equation}
    F(\mathbf{k})=\left(\mathbf{k}^{\mathsf{T}}\epsilon\mathbf{k}\right)\left(\mathbf{k}^{\mathsf{T}}\mu\mathbf{k}\right)-\left(\mathbf{k}^{\mathsf{T}}\gamma\mathbf{k}\right)^2.
\end{equation}

This single polynomial plays two roles. First, $F(\mathbf{k})=0$ is the exact condition for this zero-frequency mode coalescence. Second, its angular component $F(\hat{\mathbf{k}})$ determines whether an asymptotic propagation channel is available. If $F(\hat{\mathbf{k}})$ is sign-definite, all EFSs are closed; if it is indefinite, at least one open sheet emerges, indicating  a hyperbolic metamaterial \cite{smithElectromagneticWavePropagation2003,poddubnyHyperbolicMetamaterials2013,sunIndefiniteNatureUltraviolet2014,krishnamoorthyTopologicalTransitionsMetamaterials2012a}. The optical Lifshitz transition, corresponding to the opening or reentrant closure of these asymptotic channels, therefore occurs when $\min_{\hat{\mathbf{k}}} F(\hat{\mathbf{k}})=0 \;\text{or}\; \max_{\hat{\mathbf{k}}} F(\hat{\mathbf{k}})=0$. As illustrated in Figs.~\ref{fig:DP_EP}(b1) and \ref{fig:DP_EP}(b2), all EFSs are closed ellipsoids before the transition [Fig.~\ref{fig:DP_EP}(b1)]. Upon crossing the critical boundary, $F(\hat{\mathbf{k}})$ becomes indefinite, thereby driving the optical Lifshitz transition.

The optical Lifshitz transition boundary is closely related to the energy-metric singularity $\det(H_M)=0$, which marks the quasi-Hermitian/non-Hermitian boundary \cite{smolyaninovMetricSignatureTransitions2010,reyes-gomezMetricsignatureTopologicalTransitions2014}. For a reciprocal medium with positive-definite $\mu$ and symmetric $\gamma$, the relation is controlled by
\begin{equation}
    F(\mathbf{k}) \ge (\mathbf{k}^T\mu\mathbf{k})\left[\mathbf{k}^T(\epsilon - \gamma\mu^{-1}\gamma^T)\mathbf{k}\right].
    \label{ineq:OLT_sing_ineq}
\end{equation}
This inequality shows that metric indefiniteness is necessary for the emerengence of zero-frequency EPs. Whether an optical Lifshitz boundary coincides with $\det(H_M)=0$ is then controlled by the equality condition in Eq. \eqref{ineq:OLT_sing_ineq}, namely $\gamma \mathbf{k} = c\mu \mathbf{k}$. When this alignment condition is not satisfied, the optical opening or reclosure and the metric singularity occur at distinct boundaries.

\textit{Criticality Separation and Phase Classification:}
Table~\ref{tab:transitions} summarizes the corresponding critical conditions. Our general theory provides a systematic framework for exploring a broader constitutive-parameter space and reveals that the boundaries associated with topological transitions, EP transitions, pinch-off Lifshitz transitions, optical Lifshitz transitions, and energy-metric singularities are generally distinct. Because they are governed by different algebraic constraints, these criticalities can be explicitly decoupled through chirality and gyroelectricity. This rigorous separation opens a substantially richer parameter space, enabling the emergence of novel electromagnetic phases with exotic properties. 

To demonstrate this systematic classification, we apply our formulation to two representative models. First, as a consistency check, we revisit a uniaxially anisotropic medium with isotropic chirality, a configuration previously studied for its topological properties \cite{gaoTopologicalPhotonicPhase2015,chengNegativeRefractionsUniaxially2006,qiuBackwardWavesMagnetoelectrically2007}. Applying our theory to this model not only recovers the known phase transitions but also reveals previously unidentified phases and critical boundaries. Specifically, we find that, after the system enters the hyperbolic topological phase, a further increase in chirality causes the open EFS to reclose. The resulting two closed EFSs then separate into distinct ellipsoids, accompanied by the emergence of two finite-frequency exceptional rings. Detailed derivations for the isotropic-chirality case are provided in the Supplemental Material \cite{SM}.

\begin{figure}
    \centering
    \includegraphics[width=1\linewidth]{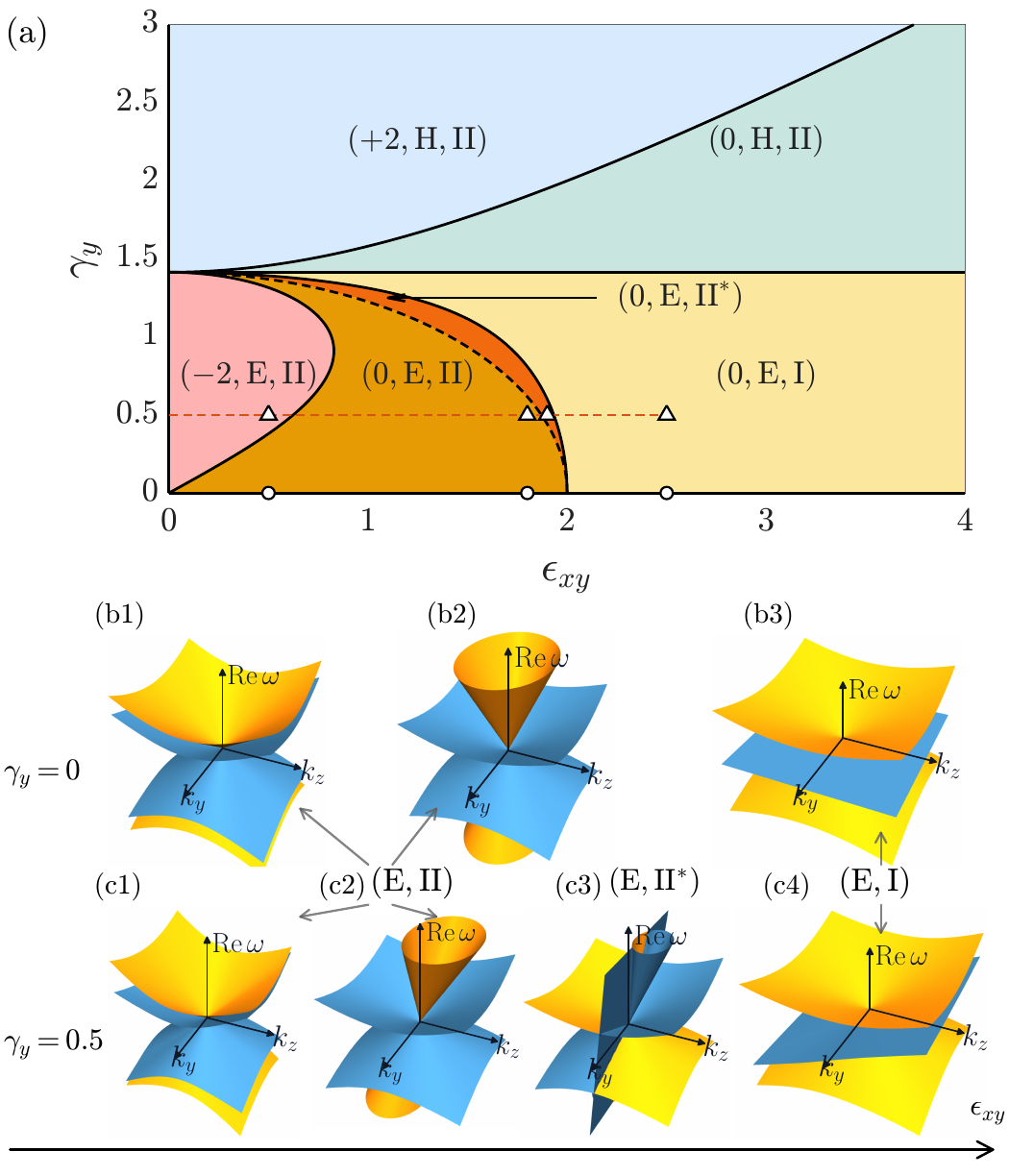}
    \caption{(a) Phase diagram of the gyroelectric-chiral medium with $\epsilon_x=2$ and $\epsilon_z=4$. Each phase is labeled in the form $(\mathcal{C}, \text{Geometry, Type})$ (see main text for detailed nomenclature). The boundary separating regions I and II marks the pinch-off Lifshitz transition. The horizontal line $\gamma_y = \sqrt{2}$ denotes the optical Lifshitz transition boundary. The dashed ellipse represents the energy metric singularity, which, in this example, also coincides with the finite-frequency EP transition. The remaining solid curves trace the two topological transition boundaries. (b, c) 2D dispersion relations $\text{Re}(\omega)$ in the $(k_y, k_z)$ plane at $k_x=0$. After filtering out the two zero-frequency modes, the yellow surfaces represent the first and fourth bands, whereas the blue surfaces represent the second and third bands. (b) Dispersion plots for $\gamma_y=0$ at $\epsilon_{xy} = 0.5, 1.8$, and $2.5$, corresponding to the parameter points marked by circles in (a). (c) Dispersion plots for $\gamma_y=0.5$ at $\epsilon_{xy} = 0.5, 1.8, 1.9$, and $2.5$, corresponding to the parameter points marked by triangles along the orange dashed line in (a). Panels (b1), (b2), (c1), and (c2) correspond to the ($\mathrm{E, II}$) phase; (c3) corresponds to the ($\mathrm{E, II}^*$) phase; and (b3), (c4) correspond to the ($\mathrm{E, I}$) phase.
    }
    \label{fig:chiral_gyroelectric_phase_diag}
\end{figure}

We next consider a gyroelectric medium with anisotropic chirality,
\begin{equation}
\epsilon=\begin{pmatrix}
\epsilon_x & i\epsilon_{xy} & 0\\
-i\epsilon_{xy} & \epsilon_x & 0\\
0 & 0 & \epsilon_z
\end{pmatrix},\;
\mu=I_{3\times 3},\;
\gamma=\mathrm{diag}(0,\gamma_y,0),
\label{eq:gyro_main_model}
\end{equation}
with $\epsilon_x=2$ and $\epsilon_z=4$. Because $F(\hat{\mathbf{k}})$ depends exclusively on the symmetric parts of the constitutive tensors $\epsilon,\mu,\gamma$, the optical Lifshitz transition boundary is independent of the antisymmetric gyroelectric parameter $\epsilon_{xy}$, reducing to the condition $\gamma_y^2 = 2$. In contrast, the energy metric becomes singular along the ellipse $\epsilon_{xy}^2 + 2\gamma_y^2 = 4$. This separation of criticalities yields the rich phase diagram in Fig.~\ref{fig:chiral_gyroelectric_phase_diag}(a).

Phases are identified by the composite label $(\mathcal{C}, \text{Geometry-Type})$, where $\mathcal{C}$ denotes the Chern number of the highest band, and the second entry specifies the EFS geometry (E/H for elliptical (closed)/hyperbolic (open), and I/II for one-sheeted/two-sheeted). ($\mathrm{E, II}^*$) designates a special ($\mathrm{E, II}$) phase where the origin is geometrically excluded from the interior of the EFS. The Chern number $\mathcal{C}$ is computed on a closed surface $S^2$ enclosing the origin of the highest band, which contains the triply degenerate point \cite{houTopologicalBandsTriply2020}: $\mathcal{C}=\frac{1}{2\pi}\oint\mathrm{d}\mathbf{k}\cdot\nabla_\mathbf{k}\times\mathbf{A}_{RR}$, where $\mathbf{A}_{RR}=-i\psi_R^\dagger\nabla_\mathbf{k}\psi_R$ is the Berry connection of the right eigenmode $\psi_R$ satisfying $H(\mathbf{k})\psi_R=\omega\psi_R$ \cite{shenTopologicalBandTheory2018,silveirinhaChernInvariantsContinuous2015,silveirinhaBulkedgeCorrespondenceTopological2016,silveirinhaTopologicalClassificationCherntype2018}.

To examine these transitions, Figs.~\ref{fig:chiral_gyroelectric_phase_diag}(b) and \ref{fig:chiral_gyroelectric_phase_diag}(c) show the dispersion relations $\text{Re}(\omega)$ evaluated at $k_x=0$. When $\gamma_y=0$ [Fig.~\ref{fig:chiral_gyroelectric_phase_diag}(b)], the system bypasses the ($\mathrm{E, II}^*$) phase. The outer yellow branch contracts symmetrically [Fig.~\ref{fig:chiral_gyroelectric_phase_diag}(b1, b2)] and pinches off at $\mathbf{k}=0$, entering the purely imaginary domain [Fig.~\ref{fig:chiral_gyroelectric_phase_diag}(b3)]. This transition to $\text{Re}(\omega)=0$ reorders the sorted spectrum, shifting this branch to the inner blue flat plane, while the remaining real-frequency branch becomes the outer yellow branch.

When $\gamma_y \neq 0$ [Fig.~\ref{fig:chiral_gyroelectric_phase_diag}(c)], the coupling between gyroelectricity and chirality breaks the $z$-axis mirror symmetry, introducing an odd-in-$k_z$ nonreciprocal term $8\epsilon_{xy}\gamma_y\omega^3 k_z$ into the dispersion equation $f(\omega, \mathbf{k})=0$, thereby shifting the EFS center \cite{SM}. The resulting non-vanishing gradient at the origin ($\partial_{k_z} f |_{\mathbf{k}=0} \neq 0$) violates the stationary condition required for the pinch-off Lifshitz transition at $\mathbf{k}=0$. At a fixed $\gamma_y=0.5$, increasing $\epsilon_{xy}$ causes the outer branch to contract and drift off-center [Fig.~\ref{fig:chiral_gyroelectric_phase_diag}(c1)]. Upon entering the ($\mathrm{E, II}^*$) phase [Fig.~\ref{fig:chiral_gyroelectric_phase_diag}(c3)], the dispersion cone tilts such that the $z$-axis is excluded from its interior except at the origin, satisfying the geometric condition to host finite-frequency exceptional rings. Finally, this pocket contracts and its eigenfrequency becomes complex [Fig.~\ref{fig:chiral_gyroelectric_phase_diag}(c4)]. Due to the reduction of its real part, this complex branch is reordered into the inner blue modes, while the remaining real branch turns yellow.

\begin{figure}
    \centering
    \includegraphics[width=\linewidth]{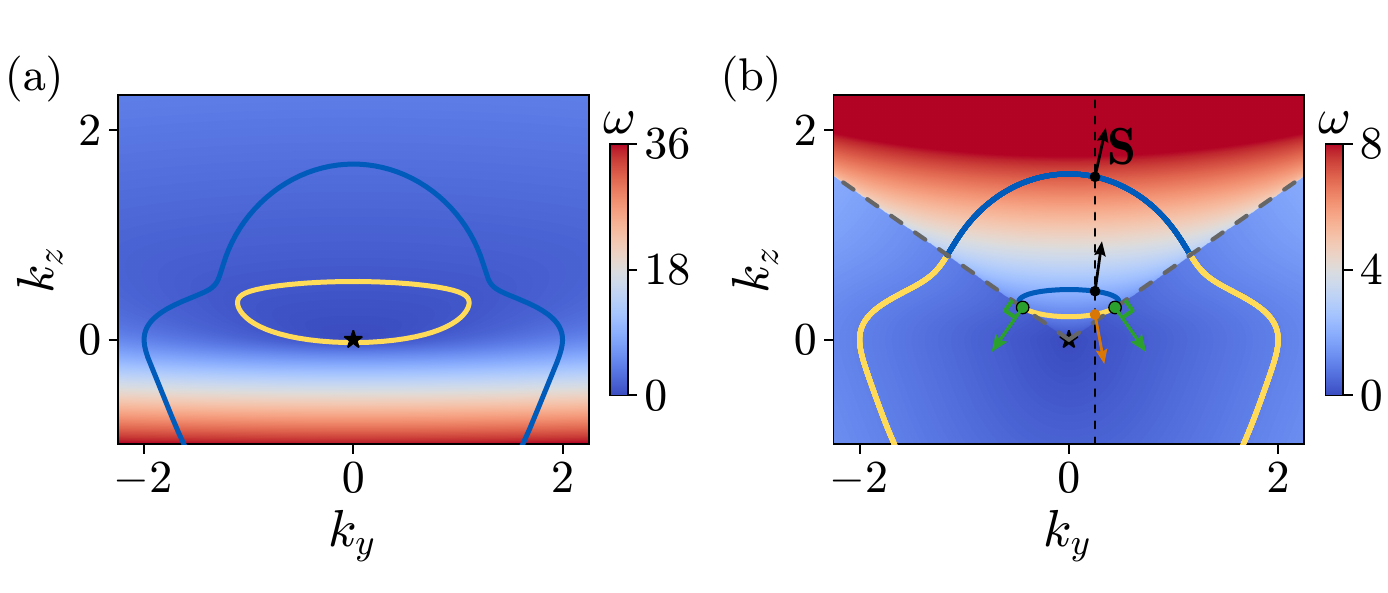}
    \caption{EFCs at $\omega=1$ and $k_x=0$ in the $(k_y, k_z)$ plane for (a) $\epsilon_{xy} = 0.75$, $\gamma_y = 1.3$, ($\mathrm{E, II}$) phase and (b) $\epsilon_{xy} = 1$, $\gamma_y = 1.3$, ($\mathrm{E, II}^*$) phase. The $\omega=1$ contours are color-coded yellow and blue to denote segments originating from the highest and non-highest bands, respectively, and the black star marks the origin. The background colormap represents the frequency of the highest band, increasing from blue to red. In (b), the arrows at the intersections of the EFCs with the black dashed line ($k_y=0.25$) indicate Poynting vector ($\mathbf{S}$) directions, where the orange intersection point marks a negative-refraction state. The gray dashed lines are tangents to the EFC and pass through the origin, where the tangent points identify two EPs (green dots), at which the Poynting vector is perpendicular to the tangent lines.}
    \label{fig:gyroelectric_EFCs}
\end{figure}

Within the ($\mathrm{E, II}^*$) phase, the exclusion of the origin from the EFC interior establishes an unconventional regime. As shown in Fig.~\ref{fig:gyroelectric_EFCs}(b), the inner closed EFC loop is divided by two EPs (green dots) into two coalescing branches. The inner branch closer to the origin belongs to the highest band (yellow segment), whereas the outer branch of the same loop belongs to a lower band (blue segment). 
These two branches coalesce at the green vertices under the geometric tangency condition $\mathbf{k} \cdot \nabla_{\mathbf{k}}f = 0$. 

These EPs mark the critical thresholds of negative refraction. Along the yellow inner branch, the Poynting vector $\mathbf{S}$, directed along the outward normal to the EFC, forms an obtuse angle with the wavevector $\mathbf{k}$, which points from the origin to the contour, yielding $\mathbf{k} \cdot \mathbf{S} < 0$. This backward-wave behavior manifests as negative refraction, with the EPs defining the critical boundary at which $\mathbf{k} \cdot \mathbf{S} = 0$.  The outer EFC loop does not participate in this coalescence; its color transition from yellow to blue is a consequence of the spectral reordering of $\text{Re}(\omega)$. This localization of negative-refraction states and the emergence of finite-frequency EPs are direct geometric consequences of the EFC displacement induced by the interplay between gyroelectricity and chirality.

\textit{Conclusion and Outlook:}
In conclusion, we have developed a general theory that establishes exact critical conditions for topological transitions, EP transitions, pinch-off Lifshitz transitions, and optical Lifshitz transitions. Our central finding is that these boundaries, often assumed to coincide, are generically separated by symmetry-breaking effects such as chirality and gyroelectricity. The resulting framework reveals a substantially richer landscape of electromagnetic phases and explicitly demonstrates how gyroelectric–chiral coupling can generate finite-frequency exceptional rings and negative refraction from a closed EFS. By systematically classifying the constitutive-parameter space and elucidating the underlying algebraic constraints, our theory provides direct guidance for designing homogenizable metamaterials with tailored electromagnetic responses.

This top-down formulation lays the foundation for several promising research directions. The framework can be readily extended to other homogenizable metamaterial systems, including magnetic and Tellegen media, to map their phase diagrams and uncover previously unexplored phenomena. Extending the theory to realistic dispersive media is another essential step toward practical device design. In addition, the intrinsic zero-frequency degeneracies supported by highly symmetric hyperbolic metamaterials may provide a natural platform for investigating higher-order EP physics. More broadly, this analytical framework enables the systematic discovery and design of novel phases in engineered electromagnetic media.

\begin{acknowledgments}
\textit{Acknowledgments:} J.Y., X.L. and C.Z. acknowledge the support of the National Science Foundation under Grant No. ECCS-2411394. Q.G. acknowledges support from the National Science Foundation ECCS-2240448.
\end{acknowledgments}

% Create the reference section using BibTeX:
\bibliography{ref}

\end{document}

\title{Supplementary Material for "Analytic Theory of Phase Transitions in Optical Metamaterials"}

\author{Jiahao Yan}
% \email{jiahao.yan@wustl.edu}
\affiliation{Department of Physics, Washington University in St. Louis, St. Louis, MO 63130, USA}
\author{Xiaoshui Lin}
\affiliation{Department of Physics, Washington University in St. Louis, St. Louis, MO 63130, USA}
\author{Junpeng Hou}
% \email{jhou@pinterest.com}
\affiliation{Pinterest Inc., San Francisco, California 94103, USA}

\author{Qing Gu}
\affiliation{Department of Electrical and Computer
Engineering, North Carolina State University, Raleigh, North
Carolina 27695, USA}
\affiliation{Department of Physics and Astronomy,
North Carolina State University, Raleigh, North Carolina
27695, USA}

\author{Chuanwei Zhang}
\email{chuanwei.zhang@wustl.edu}
\affiliation{Department of Physics, Washington University in St. Louis, St. Louis, MO 63130, USA}
\affiliation{Center for Quantum Leap, Washington University in St. Louis, St. Louis, MO 63130, USA}

\maketitle

\tableofcontents

\section{Derivation and Discussion of the Conditions of Topological Transition, EP Transition and Pinch-off Lifshitz Transition}

We begin by recalling the master equation and the associated dispersion polynomial from the generalized eigenvalue problem $H_{P}\psi=\omega H_{M}\psi$ introduced in the main text. By eliminating the magnetic field $\mathbf{H}$ from the generalized eigenvalue problem, we obtain the master equation governing the electric field $\mathbf{E}$:
\begin{eqnarray}
    &\bigg[\left(\mathbf{k}\times\right)\mu^{-1}\left(\mathbf{k}\times\right)+i\omega\left((\mathbf{k}\times)\mu^{-1}\gamma^{\mathsf{T}}+\gamma\mu^{-1}(\mathbf{k}\times)\right)\nonumber \\
    &+\omega^{2}\left(\epsilon-\gamma\mu^{-1}\gamma^{\mathsf{T}}\right)\bigg]\mathbf{E}:=K\left(\omega,\mathbf{k}\right)\mathbf{E}=0.
    \label{sup_eq:master_eq}
\end{eqnarray}
Nontrivial solutions to the master 
equation Eq. \eqref{sup_eq:master_eq} exist when $g\left(\omega,\mathbf{k}\right)\equiv\det\left(K\right)=0$, which explicitly defines the dispersion relation. The dispersion equation for a non-dispersive medium becomes a homogeneous function of $\omega$ and $k=|\mathbf{k}|$ with two static modes:
\begin{equation}
    g\left(\omega,\mathbf{k}\right)=\omega^{2}\sum_{n=0}^{4}a_{n}\left(k_{\theta},k_{\phi}\right)k^{n}\omega^{4-n}\equiv\omega^2f\left(\omega,\mathbf{k}\right)=0,
    \label{eq:homogeneous_dispersion_supp}
\end{equation}
where $k_{\theta}=\arccos \left(\frac{k_z}{k}\right)$ and $k_{\phi}= \arctan \left(\frac{k_y}{k_x}\right)$.

For the reciprocal non-dissipative medium, where $\epsilon,\mu$ are real-symmetric and $\gamma$ is real, a charge-conjugation-like symmetry $\mathcal{C}=H_M^{-1}\sigma_z\otimes I_{3\times3}T$ ensures that eigenvalues appear in $\pm\omega$ in pairs, where $T$ denotes the matrix representation of the transpose operator, i.e., $T^{-1}HT=THT^{-1}=H^{\mathsf{T}}$, and we can show that $\left\{\mathcal{C},H \right\}=0$. Then Eq.~(\ref{eq:homogeneous_dispersion_supp}) can be simplified with only even orders as:
\begin{equation} 
    f\left(\omega,\mathbf{k}\right)=\sum_{n=0}^{2}a_{2n}\left(k_{\theta},k_{\phi}\right)k^{2n}\omega^{4-2n}=0.
    \label{eq:simplified_dispersion_supp}
\end{equation}

While such symmetries permit direct analytical solutions, analyzing the critical properties of the nonlinear eigenvalue problem $K\left(\omega,\mathbf{k}\right)$ directly via its derivatives offers a more general and physically insightful framework. For a fixed wavevector $\mathbf{k}_{0}$, a root $\omega_{0}\neq0$ of $K\left(\omega,\mathbf{k}_{0}\right)$ is mathematically characterized by its algebraic and geometric multiplicities:
\\
\textit{Algebraic Multiplicity} (AM): AM$(\omega_0,\mathbf{k_0})$ is the smallest integer $m\geq1$ such that
\begin{equation}
    \partial_\omega^{(m)} \det K(\omega_0,\mathbf{k_0})\neq 0, \quad \text{while} \quad \partial_\omega^{(\ell)} \det K(\omega_0,\mathbf{k_0})=0 \quad \text{for } 0\le \ell<m.
    \label{eq:AM_supp}
\end{equation}
\textit{Geometric Multiplicity} (GM): GM$(\omega_0,\mathbf{k_0})$ is the nullity of the operator $K(\omega_0,\mathbf{k_0})$:
\begin{equation}
    \text{GM}(\omega_0, \mathbf{k_0})=\dim\ker K(\omega_0,\mathbf{k_0})=3-\mathrm{rank}\,K(\omega_0,\mathbf{k_0}).
    \label{eq:GM_supp}
\end{equation}
The conditions for band crossings and EPs follow from these definitions.
The band crossing corresponds to the degeneracy of two distinct eigenmodes, which requires AM$(\omega_0,\mathbf{k_0})=$ GM$(\omega_0,\mathbf{k_0})=2$. This implies that $\mathrm{rank}\,K(\omega_0, \mathbf{k_0})=1$. Using Jacobi's formula for the derivative of $f$ with respect to a parameter $\xi$:
\begin{equation}
    \frac{\partial f}{\partial \xi}
    =\begin{cases}
    \alpha\,\phi_{0}^{\dagger}(\partial_{\xi}K)\psi_{0} & \mathrm{rank}\left(K\right)=2, \\[2pt]
     0 & \mathrm{rank}\left(K\right)\leq1,
    \end{cases}
    \label{eq:Jacobi_supp}
\end{equation}
where $\psi_{0}$ and $\phi_{0}$ are the right and left null vectors of $K$, respectively, and $\alpha\neq0$ is a constant. The rank-1 condition for a band crossing directly implies that all first-order derivatives of $f$ must vanish. This leads to the necessary conditions $f=0$, $\partial_\omega f=0$, and $\nabla_\mathbf{k}f=\mathbf{0}$. The condition $f(\omega_0,\mathbf{k}_0)=0$ and $\nabla_\mathbf{k}f(\omega_0,\mathbf{k}_0)=\mathbf{0}$ for a given $\omega_0$ describes the touching of EFSs in $\mathbf{k}$ space. Concurrently, a band crossing is the necessary condition for a topological transition. Thus, this EFS crossing condition is also the necessary condition for a topological transition in a non-dispersive medium.

In contrast, an EP occurs when eigenmodes coalesce, resulting in a deficient eigenspace where AM$(\omega_0,\mathbf{k_0})=2$ but GM$(\omega_0,\mathbf{k_0})=1$. This occurs when $f=0$ and $\partial_\omega f=0$, but usually $\nabla_\mathbf{k}f\neq\mathbf{0}$ .

We now further explore the cases governed by the condition $\nabla_\mathbf{k}f(\omega_0,\mathbf{k}_0)=\mathbf{0}$. We expand $f(\omega_0,\mathbf{k})$ around such a point:
\begin{equation}
    f(\omega_0,\mathbf{k}) = f(\omega_0,\mathbf{k_0}) + \frac{1}{2}\mathbf{q}^\mathsf{T} f_{\mathbf{k}\mathbf{k}} \mathbf{q}+O(|\mathbf{q}|^3),
\end{equation}
where $f_{\mathbf{k}\mathbf{k}}$ is the Hessian matrix of $f$ with respect to $\mathbf{k}$, and $\mathbf{q}=\mathbf{k}-\mathbf{k}_0$.
If $f_{\mathbf{k}\mathbf{k}}$ is positive or negative definite, by Morse theory, there exists a local diffeomorphism $\mathbf{u}=\phi(\mathbf{k})$ such that
\begin{equation}
    f(\omega_0,\mathbf{u})=f(\omega_0,\mathbf{k_0})\pm(u_1^2+u_2^2+u_3^2).
\end{equation}
Consequently, when $f(\omega_0,\mathbf{k}_0)=0$, the zero set $f(\omega_0,\mathbf{k})=0$ in a neighborhood of $\mathbf{k}_0$ consists of only the single point $\mathbf{k}_0$. In the EFS picture, this implies that one branch of the EFS pinches to the point $\mathbf{k}_0$ and vanishes. We identify this pinch-off as a Lifshitz transition of the “EFS existence” type (pocket annihilation/creation), where a EFS disappears (or appears) through a pointlike collapse.

If $f_{\mathbf{k}\mathbf{k}}$ is indefinite, the function $f(\omega_0,\mathbf{k})$ can be locally mapped to the form:
\begin{equation}
    f(\omega_0,\mathbf{u})=f(\omega_0,\mathbf{k_0})+(u_1^2+u_2^2-u_3^2).
\end{equation}
This corresponds to the case where two EFS sheets intersect conically at $\mathbf{k}_0$, which is the signature of a band crossing and thus confirms the condition for a topological transition.

If the Hessian matrix $f_{\mathbf{k}\mathbf{k}}$ is degenerate, i.e., $\det(f_{\mathbf{k}\mathbf{k}})=0$, the situation is more subtle and usually corresponds to EPs. A prominent example occurs at the high-symmetry point $\mathbf{k}_0=\mathbf{0}$. At this point, $g(\omega,\mathbf{0})=\omega^6\det(\epsilon-\gamma\mu^{-1}\gamma^\mathsf{T})$. The condition $\nabla_\mathbf{k}f=\mathbf{0}$ is satisfied if $\det(\epsilon-\gamma\mu^{-1}\gamma^\mathsf{T})=0$. 
This can lead to a higher-order degeneracy because the relevant eigenvectors lie in $\ker(H_M)$
, whose dimension satisfies $\dim \ker(H_M)\leq3$, which manifests as a zero-wavevector EP.

\section{Optical Lifshitz Transition Determinant Derivation}
In this section, we provide a rigorous derivation of the determinant $F(\mathbf{k})$ introduced in the main text:
\begin{equation}
F(\mathbf{k})=\left(\mathbf{k}^\mathsf{T}\epsilon\mathbf{k}\right)\left(\mathbf{k}^\mathsf{T}\mu\mathbf{k}\right)-\left(\mathbf{k}^\mathsf{T}\gamma\mathbf{k}\right)^2.
\label{sup_eq:Fk}
\end{equation}

To establish the criteria for the optical Lifshitz transition, we analyze the dispersion equation $\det K(\omega, \mathbf{k}) = 0$ in the large-$k$ limit ($k \to \infty$). 
We employ an orthogonal basis transformation by defining the unit vector $\mathbf{p} = \mathbf{k}/k$ and a $3 \times 2$ matrix $Q = (\mathbf{e}_1, \mathbf{e}_2)$ spanning the transverse plane, such that $U = (Q, \mathbf{p})$ is a $3 \times 3$ orthogonal matrix. This basis satisfies $Q Q^{\mathsf{T}} = I_3 - \mathbf{p} \mathbf{p}^{\mathsf{T}}$, $(\mathbf{k}\times)\mathbf{p} = \mathbf{0}$, and $Q^{\mathsf{T}}(\mathbf{k}\times)Q = k J$, where $J = \left( \begin{smallmatrix} 0 & -1 \\ 1 & 0 \end{smallmatrix} \right)$ with $J^2=-I_2$.  Performing the unitary transformation $U^{\mathsf{T}}KU$, and we partition it into a $2\times2$ block form:
\begin{equation}
U^{\mathsf{T}} K U = \left( \begin{array}{cc} Q^{\mathsf{T}} K Q & Q^{\mathsf{T}} K \mathbf{p} \\ \mathbf{p}^{\mathsf{T}} K Q & \mathbf{p}^{\mathsf{T}} K \mathbf{p} \end{array} \right) = \left( \begin{array}{cc} A & B \\ C & D \end{array} \right)
\end{equation}
 We define the leading-order coefficients $A_0, B_0, C_0, D_0$ by expanding each block in powers of $k$ to identify the leading-order terms:
\begin{align}
A_0 &= Q^{\mathsf{T}}(\mathbf{k}\times)\mu^{-1}(\mathbf{k}\times)Q \quad & (O(k^2)), \\
B_0 &= Q^{\mathsf{T}}(\mathbf{k}\times)\mu^{-1}\gamma^{\mathsf{T}}\mathbf{p} \quad & (O(k)), \\
C_0 &= \mathbf{p}^{\mathsf{T}}\gamma\mu^{-1}(\mathbf{k}\times)Q \quad & (O(k)), \\
D_0 &= \mathbf{p}^{\mathsf{T}}(\epsilon - \gamma\mu^{-1}\gamma^{\mathsf{T}})\mathbf{p} \quad & (O(1)).
\end{align}
Applying the Schur complement formula $\det K = \det (U^{\mathsf{T}}KU) = \det(A) \det(D - C A^{-1} B)$ and retaining only the highest-order terms in $k$, the leading order behavior is given by
\begin{align}
    \det K &= \omega^2 \det(A_0) ( D_0 + C_0 A_0^{-1} B_0 ) + O(k^3).
    \label{sup_eq:detK}
\end{align}
First, we calculate $\det(A_0)$. Given that $(\mathbf{k}\times)Q = kQJ$, it follows that $A_0 = (k J Q^{\mathsf{T}}) \mu^{-1} (k Q J) = k^2 J (Q^{\mathsf{T}} \mu^{-1} Q) J$. The determinant is
\begin{equation}
\det(A_0) = \det(k^2 J (Q^{\mathsf{T}} \mu^{-1} Q) J) = k^4 \det(J)^2 \det(Q^{\mathsf{T}} \mu^{-1} Q) = k^4 \det(Q^{\mathsf{T}} \mu^{-1} Q).
\end{equation}
To evaluate $\det(Q^{\mathsf{T}} \mu^{-1} Q)$, we define $M = U^{\mathsf{T}} \mu^{-1} U$. By Jacobi's identity, we obtain $\det(Q^{\mathsf{T}} \mu^{-1} Q) = \det(M_{11}) = \text{cof}(M)_{33}$.
Using the identity $\text{cof}(A) = \det(A) (A^{-1})^{\mathsf{T}}$, we have
\begin{equation}
    \text{cof}(M)_{33} = (U \mathbf{e}_3)^{\mathsf{T}} \text{cof}(\mu^{-1}) (U \mathbf{e}_3) = \mathbf{p}^{\mathsf{T}} \text{cof}(\mu^{-1}) \mathbf{p};
\end{equation}
\begin{equation}
    \text{cof}(\mu^{-1}) = \det(\mu^{-1}) \mu^{\mathsf{T}}.
\end{equation}
These yield
\begin{equation}
\det(Q^{\mathsf{T}} \mu^{-1} Q) = \mathbf{p}^{\mathsf{T}} (\det(\mu^{-1}) \mu^{\mathsf{T}}) \mathbf{p} = \det(\mu^{-1}) (\mathbf{p}^{\mathsf{T}} \mu^{\mathsf{T}} \mathbf{p})= (\mathbf{p}^{\mathsf{T}} \mu\mathbf{p})/\det(\mu).
\end{equation}
Thus, we have
\begin{equation}
\det(A_0) = k^4 (\mathbf{p}^{\mathsf{T}} \mu \mathbf{p})/\det(\mu).
\label{sup_eq:detA0}
\end{equation}
Next, we calculate $C_0 A_0^{-1} B_0$. The inverse is $A_0^{-1} = [k^2 J (Q^{\mathsf{T}} \mu^{-1} Q) J]^{-1} = k^{-2} J (Q^{\mathsf{T}} \mu^{-1} Q)^{-1} J$. The full product, using $C_0 = \mathbf{p}^{\mathsf{T}}\gamma\mu^{-1}(kQJ)$ and $B_0 = (kJQ^{\mathsf{T}})\mu^{-1}\gamma^{\mathsf{T}}\mathbf{p}$, becomes
\begin{equation}
C_0 A_0^{-1} B_0 = (\mathbf{p}^{\mathsf{T}} \gamma \mu^{-1} Q) (Q^{\mathsf{T}} \mu^{-1} Q)^{-1} (Q^{\mathsf{T}} \mu^{-1} \gamma^{\mathsf{T}} \mathbf{p}).
\label{sup_eq:C0_A0_B0}
\end{equation}
To proceed, we find $(Q^{\mathsf{T}} \mu^{-1} Q)^{-1}$ by examining the blocks of $M^{-1}$
\begin{equation}
M^{-1} = U^{\mathsf{T}} \mu U = \left( \begin{array}{cc} (M^{-1})_{11} & (M^{-1})_{12} \\ (M^{-1})_{21} & (M^{-1})_{22} \end{array} \right) = \left( \begin{array}{cc} Q^{\mathsf{T}}\mu Q & Q^{\mathsf{T}}\mu \mathbf{p} \\ \mathbf{p}^{\mathsf{T}}\mu Q & \mathbf{p}^{\mathsf{T}}\mu \mathbf{p} \end{array} \right).
\end{equation}
Using the block matrix inversion formula, $(M_{11})^{-1} = (M^{-1})_{11} - (M^{-1})_{12} (M^{-1})_{22}^{-1} (M^{-1})_{21}$, we obtain the identity
\begin{equation}
(Q^{\mathsf{T}} \mu^{-1} Q)^{-1} = (Q^{\mathsf{T}} \mu Q) - (Q^{\mathsf{T}} \mu \mathbf{p}) (\mathbf{p}^{\mathsf{T}} \mu \mathbf{p})^{-1} (\mathbf{p}^{\mathsf{T}} \mu Q).
\end{equation}
Substituting this back into the expression for $C_0 A_0^{-1} B_0$ in Eq. \eqref{sup_eq:C0_A0_B0}, we find
\begin{align}
C_0 A_0^{-1} B_0 &= (\mathbf{p}^{\mathsf{T}} \gamma \mu^{-1} Q) \left[ (Q^{\mathsf{T}} \mu Q) - (Q^{\mathsf{T}} \mu \mathbf{p}) (\mathbf{p}^{\mathsf{T}} \mu \mathbf{p})^{-1} (\mathbf{p}^{\mathsf{T}} \mu Q) \right] (Q^{\mathsf{T}} \mu^{-1} \gamma^{\mathsf{T}} \mathbf{p}) \\
&= (\mathbf{p}^{\mathsf{T}} \gamma \mu^{-1} Q Q^{\mathsf{T}} \mu Q Q^{\mathsf{T}} \mu^{-1} \gamma^{\mathsf{T}} \mathbf{p}) - \frac{(\mathbf{p}^{\mathsf{T}} \gamma \mu^{-1} Q Q^{\mathsf{T}} \mu \mathbf{p}) (\mathbf{p}^{\mathsf{T}} \mu Q Q^{\mathsf{T}} \mu^{-1} \gamma^{\mathsf{T}} \mathbf{p})}{\mathbf{p}^{\mathsf{T}} \mu \mathbf{p}}.
\end{align}
Using the projector $Q Q^{\mathsf{T}} = I - \mathbf{p}\mathbf{p}^{\mathsf{T}}$, this expression can be simplified. The first term expands to
\begin{equation}
(\mathbf{p}^{\mathsf{T}} \gamma \mu^{-1} \gamma^{\mathsf{T}} \mathbf{p}) - (\mathbf{p}^{\mathsf{T}} \gamma \mathbf{p}) (\mathbf{p}^{\mathsf{T}} \mu^{-1} \gamma^{\mathsf{T}} \mathbf{p}) - (\mathbf{p}^{\mathsf{T}} \gamma \mu^{-1} \mathbf{p}) (\mathbf{p}^{\mathsf{T}} \gamma^{\mathsf{T}} \mathbf{p}) + (\mathbf{p}^{\mathsf{T}} \gamma \mu^{-1} \mathbf{p}) (\mathbf{p}^{\mathsf{T}} \mu \mathbf{p}) (\mathbf{p}^{\mathsf{T}} \mu^{-1} \gamma^{\mathsf{T}} \mathbf{p}).
\end{equation}
The second term expands to
\begin{equation}
\frac{(\mathbf{p}^{\mathsf{T}} \gamma \mathbf{p})(\mathbf{p}^{\mathsf{T}} \gamma^{\mathsf{T}} \mathbf{p})}{\mathbf{p}^{\mathsf{T}} \mu \mathbf{p}} - (\mathbf{p}^{\mathsf{T}} \gamma \mathbf{p}) (\mathbf{p}^{\mathsf{T}} \mu^{-1} \gamma^{\mathsf{T}} \mathbf{p}) - (\mathbf{p}^{\mathsf{T}} \gamma \mu^{-1} \mathbf{p}) (\mathbf{p}^{\mathsf{T}} \gamma^{\mathsf{T}} \mathbf{p}) + (\mathbf{p}^{\mathsf{T}} \gamma \mu^{-1} \mathbf{p}) (\mathbf{p}^{\mathsf{T}} \mu \mathbf{p}) (\mathbf{p}^{\mathsf{T}} \mu^{-1} \gamma^{\mathsf{T}} \mathbf{p}).
\end{equation}
Subtracting the second expansion from the first, the last three parts can be canceled exactly, leaving
\begin{equation}
C_0 A_0^{-1} B_0 = (\mathbf{p}^{\mathsf{T}} \gamma \mu^{-1} \gamma^{\mathsf{T}} \mathbf{p}) - \frac{(\mathbf{p}^{\mathsf{T}} \gamma \mathbf{p}) (\mathbf{p}^{\mathsf{T}} \gamma^{\mathsf{T}} \mathbf{p})}{\mathbf{p}^{\mathsf{T}} \mu \mathbf{p}}.
\end{equation}
Finally, the scalar part $D_0 + C_0 A_0^{-1} B_0$ in Eq. \eqref{sup_eq:detK} becomes
\begin{equation}
\left[ (\mathbf{p}^{\mathsf{T}} \epsilon \mathbf{p}) - (\mathbf{p}^{\mathsf{T}} \gamma \mu^{-1} \gamma^{\mathsf{T}} \mathbf{p}) \right] + \left[ (\mathbf{p}^{\mathsf{T}} \gamma \mu^{-1} \gamma^{\mathsf{T}} \mathbf{p}) - \frac{(\mathbf{p}^{\mathsf{T}} \gamma \mathbf{p}) (\mathbf{p}^{\mathsf{T}} \gamma^{\mathsf{T}} \mathbf{p})}{\mathbf{p}^{\mathsf{T}} \mu \mathbf{p}} \right] = \frac{(\mathbf{p}^{\mathsf{T}} \epsilon \mathbf{p}) (\mathbf{p}^{\mathsf{T}} \mu \mathbf{p}) - (\mathbf{p}^{\mathsf{T}} \gamma \mathbf{p}) (\mathbf{p}^{\mathsf{T}} \gamma^{\mathsf{T}} \mathbf{p})}{\mathbf{p}^{\mathsf{T}} \mu \mathbf{p}}.
\end{equation}
Substituting this and $\det(A_0)$ in Eq. \eqref{sup_eq:detA0} into the leading-order expression for $\det K$ that we derived in Eq. \eqref{sup_eq:detK}, we have
\begin{align}
    \det K &\approx \omega^2 k^4 \left[ (\mathbf{p}^{\mathsf{T}} \epsilon \mathbf{p}) (\mathbf{p}^{\mathsf{T}} \mu \mathbf{p}) - (\mathbf{p}^{\mathsf{T}} \gamma \mathbf{p}) (\mathbf{p}^{\mathsf{T}} \gamma^{\mathsf{T}} \mathbf{p}) \right]/\det(\mu)\\
&=\omega^2  \left[ (\mathbf{k}^{\mathsf{T}} \epsilon \mathbf{k}) (\mathbf{k}^{\mathsf{T}} \mu \mathbf{k}) - (\mathbf{k}^{\mathsf{T}} \gamma \mathbf{k}) (\mathbf{k}^{\mathsf{T}} \gamma^{\mathsf{T}} \mathbf{k}) \right]/\det(\mu)\\
&=\omega^2  \left[ (\mathbf{k}^{\mathsf{T}} \epsilon \mathbf{k}) (\mathbf{k}^{\mathsf{T}} \mu \mathbf{k}) - (\mathbf{k}^{\mathsf{T}} \gamma \mathbf{k})^2 \right]/\det(\mu).
\end{align}
And we take $F(\mathbf{k})\propto\det K = (\mathbf{k}^{\mathsf{T}} \epsilon \mathbf{k}) (\mathbf{k}^{\mathsf{T}} \mu \mathbf{k}) - (\mathbf{k}^{\mathsf{T}} \gamma \mathbf{k})^2$, which is exactly the same as that in Eq. \eqref{sup_eq:Fk}.

\section{Uniaxial Medium with Isotropic Chirality}

We now apply the general criteria derived above to the uniaxial isotropic-chirality model discussed in the main text. Specializing the model to $\mu_0=1$ and $\epsilon_x=4$, we obtain
\begin{equation}
\epsilon=\mathrm{diag}(4,4,\epsilon_z),\qquad
\mu=I_{3\times 3},\qquad
\gamma=\gamma_0 I_{3\times 3}.
\label{sup_eq:isoep_model}
\end{equation}
The propagating modes are governed by
\begin{align}
f(\omega,\mathbf{k})={}&(4-\gamma_0^2)^2(\epsilon_z-\gamma_0^2)\omega^4 \nonumber\\
&-\Big[(4-\gamma_0^2)(4+\epsilon_z+2\gamma_0^2)(k_x^2+k_y^2)
+2(4+\gamma_0^2)(\epsilon_z-\gamma_0^2)k_z^2\Big]\omega^2 \nonumber\\
&+(k_x^2+k_y^2+k_z^2)\Big[(4-\gamma_0^2)(k_x^2+k_y^2)+(\epsilon_z-\gamma_0^2)k_z^2\Big]=0,
\label{sup_eq:isoep_dispersion}
\end{align}
while the large-$k$ form introduced in Sec.~II becomes
\begin{align}
F(\mathbf{k})={}&\left[4(k_x^2+k_y^2)+\epsilon_z k_z^2\right](k_x^2+k_y^2+k_z^2)
-\gamma_0^2(k_x^2+k_y^2+k_z^2)^2 \nonumber\\
={}&(k_x^2+k_y^2+k_z^2)\Big[(4-\gamma_0^2)(k_x^2+k_y^2)+(\epsilon_z-\gamma_0^2)k_z^2\Big].
\label{sup_eq:isoep_F}
\end{align}

\subsection{Optical Lifshitz transition and energy metric singularity}
Equation~\eqref{sup_eq:isoep_F} shows that the asymptotic EFS is open when $4-\gamma_0^2$ and $\epsilon_z-\gamma_0^2$ have opposite signs, and closed when they have the same sign. The EFS opening or reclosure is therefore controlled by
\begin{equation}
\gamma_0^2=4,\qquad \gamma_0^2=\epsilon_z.
\label{sup_eq:isoep_optical_boundaries}
\end{equation}
For instance, at $\epsilon_z = -6$, the medium supports an open hyperbolic EFS at $\gamma_0=0$. As the isotropic chirality increases, the asymptotic channels reclose at $\gamma_0^2=4$ (i.e., $|\gamma_0|=2$), manifesting a chirality-driven optical Lifshitz transition.

The energy metric satisfies
\begin{equation}
\det(H_M)=\det(\epsilon-\gamma\mu^{-1}\gamma^{\mathsf T})=(4-\gamma_0^2)^2(\epsilon_z-\gamma_0^2),
\label{sup_eq:isoep_metric}
\end{equation}
so, for this model, the metric singularity occurs exactly under the same critical conditions as the optical Lifshitz transition.

\subsection{Finite-frequency stationary conditions and topological phase transition candidates}
We now discuss the finite-frequency topological phase transition problem. At a prescribed $\omega\neq0$, the candidates satisfy $f=0$ together with
\begin{align}
\partial_{k_x} f={}&2k_x\Big[-(4-\gamma_0^2)(4+\epsilon_z+2\gamma_0^2)\omega^2 \nonumber\\
&\hspace{1.7cm}+2(4-\gamma_0^2)(k_x^2+k_y^2)+(4+\epsilon_z-2\gamma_0^2)k_z^2\Big], \label{sup_eq:isoep_dkx}\\
\partial_{k_y} f={}&2k_y\Big[-(4-\gamma_0^2)(4+\epsilon_z+2\gamma_0^2)\omega^2 \nonumber\\
&\hspace{1.7cm}+2(4-\gamma_0^2)(k_x^2+k_y^2)+(4+\epsilon_z-2\gamma_0^2)k_z^2\Big], \label{sup_eq:isoep_dky}\\
\partial_{k_z} f={}&2k_z\Big[-2(4+\gamma_0^2)(\epsilon_z-\gamma_0^2)\omega^2 \nonumber\\
&\hspace{1.7cm}+(4+\epsilon_z-2\gamma_0^2)(k_x^2+k_y^2)+2(\epsilon_z-\gamma_0^2)k_z^2\Big].
\label{sup_eq:isoep_dkz}
\end{align}
These equations naturally separate into three cases.

If $k_x=k_y=0$, then
\begin{equation}
f(\omega,0,0,k_z)=(\epsilon_z-\gamma_0^2)\left[k_z^2-(\gamma_0+2)^2\omega^2\right]\left[k_z^2-(\gamma_0-2)^2\omega^2\right].
\label{sup_eq:isoep_axis_factor}
\end{equation}
A double root occurs only at
\begin{equation}
\gamma_0=0,\qquad k_x=k_y=0,\qquad k_z=\pm2\omega.
\label{sup_eq:isoep_axis_family}
\end{equation}
This is the well-known topological phase transition critical point \cite{gaoTopologicalPhotonicPhase2015}.

If $k_z=0$ but $k_x^2+k_y^2\neq0$, Eqs.~\eqref{sup_eq:isoep_dkx} and \eqref{sup_eq:isoep_dky} give
\begin{equation}
k_x^2+k_y^2=\frac{4+\epsilon_z+2\gamma_0^2}{2}\,\omega^2,
\label{sup_eq:isoep_equatorial_radius}
\end{equation}
and substituting this into $f=0$ yields
\begin{equation}
(\epsilon_z-4)^2+8\gamma_0^2(\epsilon_z+4)=0.
\label{sup_eq:isoep_equatorial_boundary}
\end{equation}
Owing to the in-plane rotational symmetry, we may set $(k_x,k_y,k_z)=(\kappa,0,0)$ with $\kappa^2=(4+\epsilon_z+2\gamma_0^2)\omega^2/2$. The matrix then becomes
\begin{equation}
K(\omega,\kappa,0,0)=
\begin{pmatrix}
(4-\gamma_0^2)\omega^2 & 0 & 0\\
0 & (4-\gamma_0^2)\omega^2-\kappa^2 & -2i\gamma_0\kappa\omega\\
0 & 2i\gamma_0\kappa\omega & (\epsilon_z-\gamma_0^2)\omega^2-\kappa^2
\end{pmatrix}.
\label{sup_eq:isoep_equatorial_K}
\end{equation}
On Eq.~\eqref{sup_eq:isoep_equatorial_boundary}, the $2\times2$ block has determinant
\begin{equation}
-\frac{\omega^4}{4}\Big[(\epsilon_z-4)^2+8\gamma_0^2(\epsilon_z+4)\Big]=0,
\label{sup_eq:isoep_equatorial_det2}
\end{equation}
while its trace is $-4\gamma_0^2\omega^2\neq0$ on the physical branch. Hence the block has rank $1$, the first diagonal entry in Eq.~\eqref{sup_eq:isoep_equatorial_K} is nonzero, and therefore
\begin{equation}
\mathrm{rank}\,K=2,\qquad \mathrm{GM}=1.
\label{sup_eq:isoep_equatorial_rank}
\end{equation}
At the same time,
\begin{equation}
\partial_\omega^2 f=-8(\gamma_0^2-\epsilon_z)(\gamma_0^2-4)^2\omega^2\neq0,
\label{sup_eq:isoep_equatorial_am}
\end{equation}
so AM$=2$. This equatorial family is therefore an EP transition, not a topological phase transition.

\begin{figure}[t]
\centering
\includegraphics[width=1.0\linewidth]{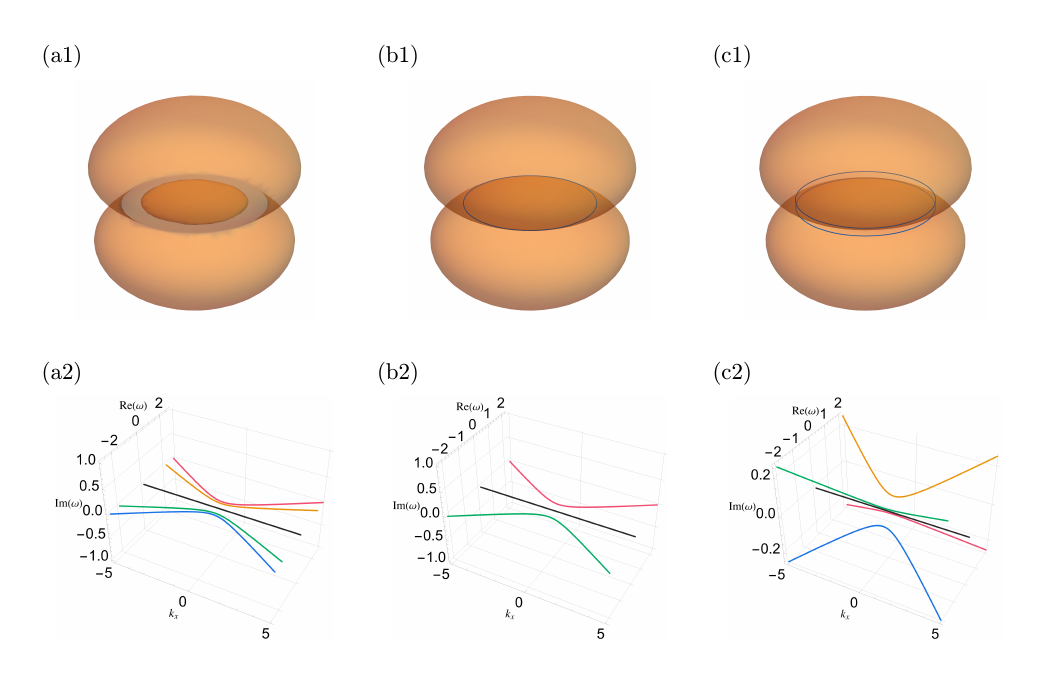}
\caption{Critical evolution across the equatorial boundary in Eq.~\eqref{sup_eq:isoep_equatorial_boundary} at $\epsilon_z=-6$. Columns (a), (b), and (c) correspond to $\gamma_0=2.4$, $2.5$, and $2.6$, respectively. The upper row shows the EFSs at $\omega=1$, and the lower row shows the band structure along the cut $k_y=1$, $k_z=0$ as a function of $k_x$. At the critical value $\gamma_0=2.5$, the two real-frequency band pairs coalesce on the equatorial critical ring; beyond the boundary they split into complex branches.}
\label{fig:equatorial_critical_EP}
\end{figure}

For $\epsilon_z=-6$, Eq.~\eqref{sup_eq:isoep_equatorial_boundary} gives the critical value $\gamma_0=2.5$. Fig.~\ref{fig:equatorial_critical_EP} shows the evolution across this boundary on the cut $k_y=1$, $k_z=0$: the red-yellow pair and the blue-green pair first coalesce at the critical ring respectively, and then split into complex branches, which is the characteristic spectral evolution across an EP. Beyond this boundary, the intersecting ring unfolds into finite-frequency exceptional rings while the asymptotic channel remains closed.

Finally, if both $k_x^2+k_y^2\neq0$ and $k_z\neq0$, Eqs.~\eqref{sup_eq:isoep_dkx}--\eqref{sup_eq:isoep_dkz} give
\begin{align}
k_x^2+k_y^2&=\frac{4\gamma_0^2(\epsilon_z-\gamma_0^2)}{\epsilon_z-4}\,\omega^2, \label{sup_eq:isoep_mixed_kxy}\\
k_z^2&=\frac{(\gamma_0^2-4)(4\gamma_0^2+4-\epsilon_z)}{\epsilon_z-4}\,\omega^2.
\label{sup_eq:isoep_mixed_kz}
\end{align}
Substituting these into Eq.~\eqref{sup_eq:isoep_dispersion} yields
\begin{equation}
4\gamma_0^2(\epsilon_z-\gamma_0^2)(\gamma_0^2-4)\omega^4=0.
\label{sup_eq:isoep_mixed_constraint}
\end{equation}
Thus there is no additional topological transition candidates away from the already identified special set $\gamma_0=0$, and ill-defined sets $\gamma_0^2=4$, and $\gamma_0^2=\epsilon_z$. Collecting the three cases above, the only topological phase transition critical condition is Eq.~\eqref{sup_eq:isoep_axis_family}. The critical boundaries identified above classify the topological phases of the system. For $\epsilon_z = -6$, the intermediate regimes between transitions ($0 < |\gamma_0| < 2$ and $2 < |\gamma_0| < 2.5$) feature a spectrum that is fully gapped on the momentum-space unit sphere. The system hosts a topological phase characterized by a Chern number $|\mathcal{C}| = 2$, computed on a closed surface $S^2$ enclosing the origin via $\mathcal{C}=\frac{1}{2\pi}\oint\mathrm{d}\mathbf{k}\cdot\nabla_\mathbf{k}\times\mathbf{A}_{RR}$, where $\mathbf{A}_{RR}=-i\psi_R^\dagger\nabla_\mathbf{k}\psi_R$ is the Berry connection of the right eigenmode. For $|\gamma_0| > 2.5$, the unfolding of the stationary ring into finite-frequency exceptional rings prevents the highest band from forming a globally isolated complex line bundle over the entire momentum sphere, rendering the single band Chern number ill-defined.

\subsection{EP transitions}
At zero frequency, the EP manifold is determined by
\begin{equation}
F(\mathbf{k})=0
\quad\Longleftrightarrow\quad
(4-\gamma_0^2)(k_x^2+k_y^2)+(\epsilon_z-\gamma_0^2)k_z^2=0.
\label{sup_eq:isoep_zero_EP}
\end{equation}
 For $\min(4,\epsilon_z)<\gamma_0^2<\max(4,\epsilon_z)$, Eq.~\eqref{sup_eq:isoep_zero_EP} describes an EP cone in $\mathbf{k}$ space.

\begin{figure}[t]
\centering
\includegraphics[width=0.98\linewidth]{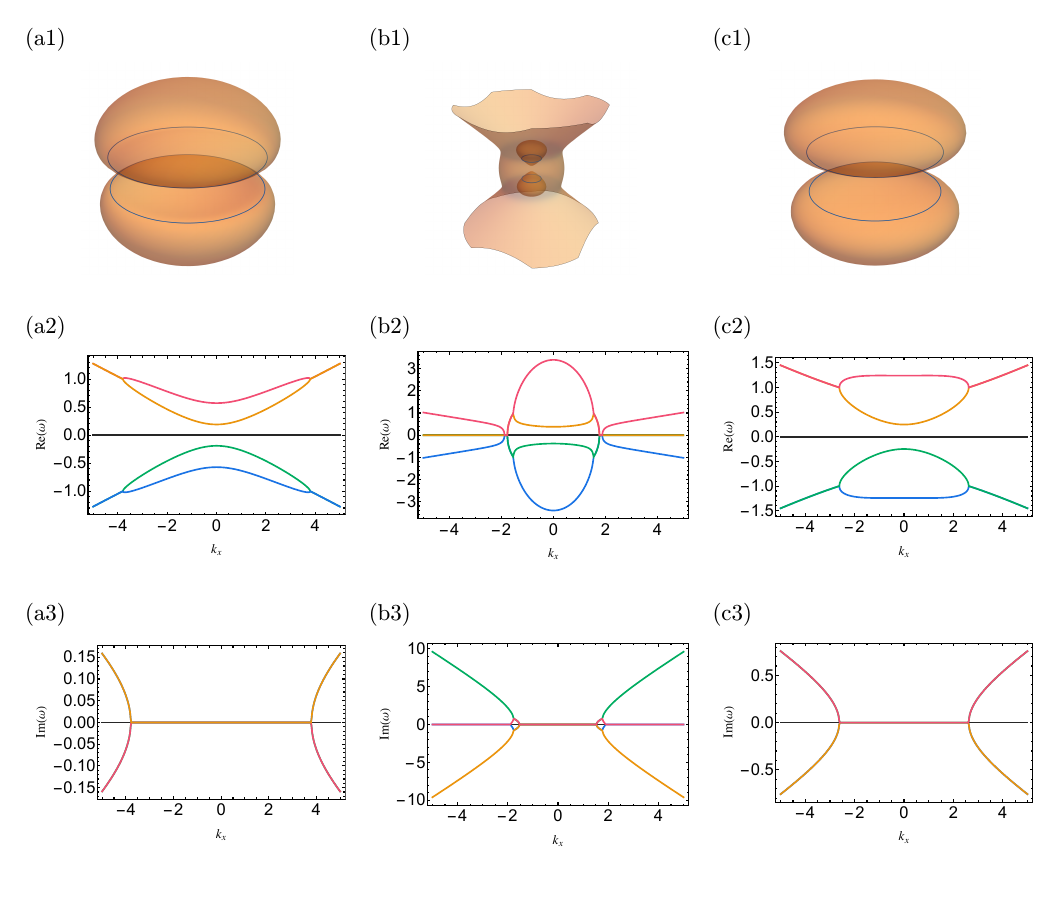}
\caption{Representative EFSs and 1d bands in the three finite-frequency EP sectors. Columns (a), (b), and (c) correspond to $(\epsilon_z,\gamma_0)=(-6,4)$, $(9,2.5)$, and $(-15,3)$, respectively. The top row shows the EFSs at $\omega=1$ with the relevant EP rings highlighted. The middle and bottom rows show $\mathrm{Re}\,\omega$ and $\mathrm{Im}\,\omega$ as functions of $k_x$ on the cut $k_y=0$, with $k_z$ chosen from the $\omega=1$ EP-ring equation. Panels (a) and (c) exhibit finite-frequency EPs at $\omega=1$. Panel (b) contains both the $\omega=0$ and $\omega=1$ EPs on the same cut; an additional purely imaginary EP is also visible there but lies outside the present discussion.}
\label{fig:EP_ring_examples}
\end{figure}

At finite frequency, EPs satisfy $f=0$ and $\partial_\omega f=0$. The physical branch at a prescribed frequency is
\begin{equation}
\begin{split}
k_z^2={}&\omega^2\frac{\gamma_0^2-4}{(\epsilon_z-4)(3\gamma_0^2+4)}
\Big[(\gamma_0^2-4)(\epsilon_z-4\gamma_0^2-4)\\
&\hspace{2.0cm}+2(\epsilon_z+2\gamma_0^2+4)\sqrt{\gamma_0^2(\gamma_0^2-4)}\Big],
\end{split}
\label{sup_eq:isoep_kz_EP}
\end{equation}
\begin{equation}
\begin{split}
k_x^2+k_y^2={}&4\omega^2\frac{\gamma_0^2-\epsilon_z}{(\epsilon_z-4)(3\gamma_0^2+4)}
\Big[\gamma_0^2(\gamma_0^2-4)\\
&\hspace{2.15cm}-(\gamma_0^2+4)\sqrt{\gamma_0^2(\gamma_0^2-4)}\Big].
\end{split}
\label{sup_eq:isoep_kxy_EP}
\end{equation}
For each fixed $\omega\neq0$, Eqs.~\eqref{sup_eq:isoep_kz_EP} and \eqref{sup_eq:isoep_kxy_EP} describe a pair of exceptional rings related by $k_z\to-k_z$, except when Eq.~\eqref{sup_eq:isoep_equatorial_boundary} is reached and the pair collapses to the equatorial critical EP ring with $k_z=0$. Because the model is homogeneous, the ring radii scale linearly with $|\omega|$.

The finite-frequency EP sector is
\begin{align}
&\epsilon_z>4,\qquad 4<\gamma_0^2<\epsilon_z, \nonumber\\
&-12<\epsilon_z<-4,\qquad (\epsilon_z-4)^2+8\gamma_0^2(\epsilon_z+4)\le0, \label{sup_eq:isoep_EP_windows}\\
&\epsilon_z\le-12,\qquad \gamma_0^2>4. \nonumber
\end{align}
In the middle line, the equality sign gives the equatorial critical EP ring in Eq.~\eqref{sup_eq:isoep_equatorial_boundary}, while the strict inequality gives the off-equatorial pair $k_z=\pm\sqrt{k_z^2}$. There is no finite-frequency EP for $-4\le\epsilon_z\le4$. For $\epsilon_z>4$, the rings are created from the origin at $\gamma_0^2=4$ and shrink onto the $k_z$ axis as $\gamma_0^2\to\epsilon_z^-$. For $-12<\epsilon_z<-4$, the equatorial critical ring marks the birth of the off-equatorial EP pair. For $\epsilon_z\le-12$, the rings appear once $\gamma_0^2>4$.

Representative examples from these three finite-frequency EP sectors are shown in Fig.~\ref{fig:EP_ring_examples}.

Here is a brief summary of several EP sectors:
\begin{itemize}
\item $\epsilon_z>4$: zero-frequency EP cone for $4\le\gamma_0^2\le\epsilon_z$, and finite-frequency EP rings for $4<\gamma_0^2<\epsilon_z$.
\item $0\le\epsilon_z\le4$: zero-frequency EP cone for $\epsilon_z\le\gamma_0^2\le4$.
\item $-4\le\epsilon_z<0$: zero-frequency EP cone for $0\le\gamma_0^2\le4$.
\item $-12<\epsilon_z<-4$: zero-frequency EP cone for $0\le\gamma_0^2\le4$; the equatorial line in Eq.~\eqref{sup_eq:isoep_equatorial_boundary} is a finite-frequency critical EP ring; for strict inequality in Eq.~\eqref{sup_eq:isoep_EP_windows}, the EP ring splits into an off-equatorial pair.
\item $\epsilon_z\le-12$: zero-frequency EP cone for $0\le\gamma_0^2\le4$, and off-equatorial finite-frequency EP rings for $\gamma_0^2>4$.
\end{itemize}

\section{Gyroelectric Medium with Anisotropic Chirality}

We next turn to the gyroelectric-chiral model discussed in Fig.~2 and Fig.3 of the main text. Specializing Eq.~(6) of the main text to $\epsilon_x=2$ and $\epsilon_z=4$, we take
\begin{equation}
\epsilon=
\begin{pmatrix}
2 & i\epsilon_{xy} & 0\\
-i\epsilon_{xy} & 2 & 0\\
0 & 0 & 4
\end{pmatrix},
\qquad
\mu=I_{3\times3},
\qquad
\gamma=\mathrm{diag}(0,\gamma_y,0).
\label{sup_eq:gyro_model}
\end{equation}
The propagating modes are governed by
\begin{align}
f(\omega,\mathbf{k})={}&\big[2(k_x^2+k_y^2)+4k_z^2\big](k_x^2+k_y^2+k_z^2)-\gamma_y^2k_y^4 \nonumber\\
&-2\epsilon_{xy}\gamma_y k_y^2k_z\omega+8\epsilon_{xy}\gamma_y k_z\omega^3 \nonumber\\
&+\big[\epsilon_{xy}^2(k_x^2+k_y^2)+6\gamma_y^2k_y^2-12(k_x^2+k_y^2)-16k_z^2\big]\omega^2 \nonumber\\
&+4(4-\epsilon_{xy}^2-2\gamma_y^2)\omega^4=0,
\label{sup_eq:gyro_dispersion}
\end{align}
while the large-$k$ form introduced in Sec.~II becomes
\begin{align}
F(\mathbf{k})={}&(\mathbf{k}^{\mathsf T}\epsilon\mathbf{k})(\mathbf{k}^{\mathsf T}\mu\mathbf{k})-(\mathbf{k}^{\mathsf T}\gamma\mathbf{k})^2 \nonumber\\
={}&\big[2(k_x^2+k_y^2)+4k_z^2\big](k_x^2+k_y^2+k_z^2)-\gamma_y^2k_y^4.
\label{sup_eq:gyro_F}
\end{align}

\subsection{Optical Lifshitz transition, zero-frequency EP, and energy metric singularity}
\begin{figure}[t]
\centering
\includegraphics[width=0.8\linewidth]{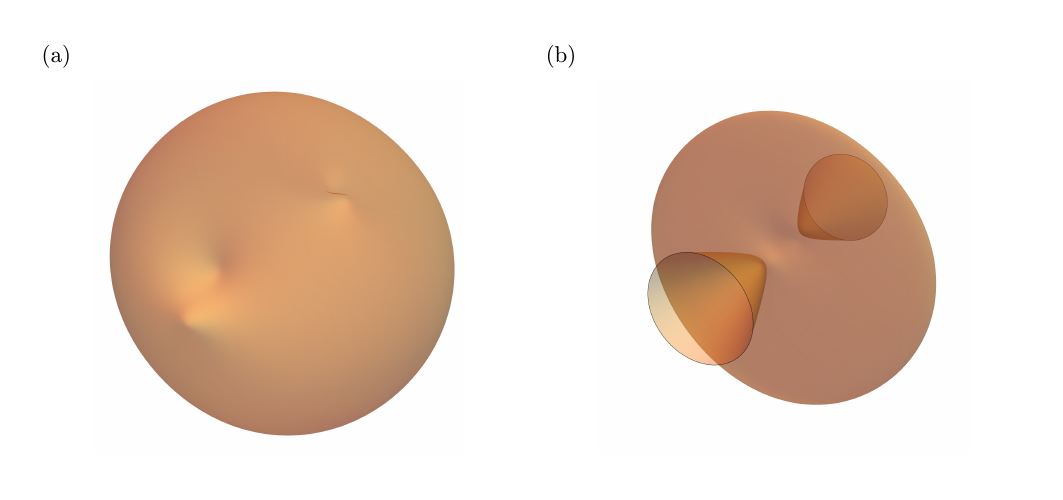}
\caption{Representative EFSs across the optical Lifshitz transition in the gyroelectric-chiral model at fixed $\epsilon_{xy}=0.5$. (a): $\gamma_y=\sqrt{2}-0.01$, for which the asymptotic EFS is closed. (b): $\gamma_y=\sqrt{2}+0.01$, for which open asymptotic channels are present. The change in asymptotic connectivity occurs at $\gamma_y^2=2$ [Eq.~\eqref{sup_eq:gyro_optical}].}
\label{fig:gyro_olt}
\end{figure}

Because the gyroelectric contribution is antisymmetric, it drops out of $\mathbf{k}^{\mathsf T}\epsilon\mathbf{k}$ and therefore does not enter Eq.~\eqref{sup_eq:gyro_F}. Writing $\hat{\mathbf{k}}=\mathbf{k}/|\mathbf{k}|$, we obtain
\begin{equation}
F(\hat{\mathbf{k}})=2+2\hat{k}_z^2-\gamma_y^2\hat{k}_y^4.
\label{sup_eq:gyro_Fhat}
\end{equation}
Its minimum is $2-\gamma_y^2$, attained at $\hat{\mathbf{k}}=\pm\hat{\mathbf{y}}$, and its maximum is $4$, attained at $\hat{\mathbf{k}}=\pm\hat{\mathbf{z}}$. Therefore the asymptotic EFS is closed for $\gamma_y^2<2$ and open for $\gamma_y^2>2$, so the optical Lifshitz transition occurs at
\begin{equation}
\gamma_y^2=2.
\label{sup_eq:gyro_optical}
\end{equation}
Since $\max_{\hat{\mathbf{k}}}F(\hat{\mathbf{k}})=4>0$ always, the zero-frequency EP condition $F(\mathbf{k})=0$ has real solutions whenever Eq.~\eqref{sup_eq:gyro_optical} is crossed, i.e., for
\begin{equation}
\gamma_y^2\ge 2.
\label{sup_eq:gyro_zero_EP}
\end{equation}
The energy metric instead satisfies
\begin{equation}
\det(H_M)=\det(\epsilon-\gamma\mu^{-1}\gamma^{\mathsf T})=4(4-\epsilon_{xy}^2-2\gamma_y^2),
\label{sup_eq:gyro_metric}
\end{equation}
so its singularity is the ellipse
\begin{equation}
\epsilon_{xy}^2+2\gamma_y^2=4.
\label{sup_eq:gyro_metric_curve}
\end{equation}
Equations~\eqref{sup_eq:gyro_optical} and \eqref{sup_eq:gyro_metric_curve} therefore coincide only at $(\epsilon_{xy},\gamma_y)=(0,\sqrt{2})$, which is the criticality separation mentioned in the main text. At fixed $\epsilon_{xy}=0.5$, representative EFSs on the two sides of Eq.~\eqref{sup_eq:gyro_optical} are shown in Fig.~\ref{fig:gyro_olt}. The surface is still closed at $\gamma_y=\sqrt{2}-0.01$, whereas open EFS appear at $\gamma_y=\sqrt{2}+0.01$, directly illustrating the optical Lifshitz transition.

\subsection{Finite-frequency stationary conditions: topological and pinch-off Lifshitz transitions}
We now discuss the finite-frequency stationary problem. At a prescribed $\omega\neq0$, the candidates satisfy $f=0$ together with
\begin{align}
\partial_{k_x}f={}&2k_x\Big[\epsilon_{xy}^2\omega^2+4(k_x^2+k_y^2)+6k_z^2-12\omega^2\Big],
\label{sup_eq:gyro_dkx}\\
\partial_{k_y}f={}&2k_y\Big[\epsilon_{xy}^2\omega^2-2\epsilon_{xy}\gamma_y k_z\omega-2\gamma_y^2k_y^2+6\gamma_y^2\omega^2 \nonumber\\
&\hspace{2.6cm}+4(k_x^2+k_y^2)+6k_z^2-12\omega^2\Big],
\label{sup_eq:gyro_dky}\\
\partial_{k_z}f={}&2\Big[-\epsilon_{xy}\gamma_y k_y^2\omega+4\epsilon_{xy}\gamma_y\omega^3+6(k_x^2+k_y^2)k_z+8k_z^3-16k_z\omega^2\Big].
\label{sup_eq:gyro_dkz}
\end{align}
The physical stationary families relevant to the phase diagram are as follows.

If $k_x=0$, $k_y\neq0$, and $k_z\neq0$, direct solution of Eqs.~\eqref{sup_eq:gyro_dky} and \eqref{sup_eq:gyro_dkz} gives
\begin{equation}
k_x=0,
\qquad
k_y=\pm\frac{2\gamma_y}{\sqrt{\gamma_y^2+2}}\,\omega,
\qquad
k_z=\frac{\sqrt{4-2\gamma_y^2}}{\sqrt{\gamma_y^2+2}}\,\omega,
\label{sup_eq:gyro_lower_topological_point}
\end{equation}
together with the parameter relation
\begin{equation}
\epsilon_{xy}^2=\frac{2\gamma_y^2(2-\gamma_y^2)}{\gamma_y^2+2},
\qquad 0\le\gamma_y^2\le2.
\label{sup_eq:gyro_lower_topological_curve}
\end{equation}
At the stationary point, the Hessian with respect to $(k_x,k_y,k_z)$ is
\begin{equation}
f_{\mathbf{k}\mathbf{k}}=\omega^2
\begin{pmatrix}
-4\gamma_y^2 & 0 & 0\\
0 & \dfrac{32\gamma_y^2(2-\gamma_y^2)}{\gamma_y^2+2} & \dfrac{8\gamma_y\sqrt{4-2\gamma_y^2}(6-\gamma_y^2)}{\gamma_y^2+2}\\[8pt]
0 & \dfrac{8\gamma_y\sqrt{4-2\gamma_y^2}(6-\gamma_y^2)}{\gamma_y^2+2} & \dfrac{16(8-5\gamma_y^2)}{\gamma_y^2+2}
\end{pmatrix},
\label{sup_eq:gyro_lower_hessian}
\end{equation}
whose lower $2\times2$ block has determinant
\begin{equation}
\det\big[(f_{\mathbf{k}\mathbf{k}})_{yz}\big]
=\frac{128\gamma_y^2(\gamma_y^2-2)(\gamma_y^4+8\gamma_y^2+4)}{(\gamma_y^2+2)^2}\,\omega^4<0
\label{sup_eq:gyro_lower_hessian_det}
\end{equation}
for $0<\gamma_y^2<2$. The Hessian is therefore indefinite, so Eq.~\eqref{sup_eq:gyro_lower_topological_curve} is a genuine topological transition boundary. It separates the closed-sheet sectors $(-2, \mathrm{E, II})$ and $(0, \mathrm{E, II})$.

\begin{figure}[t]
\centering
\includegraphics[width=0.98\linewidth]{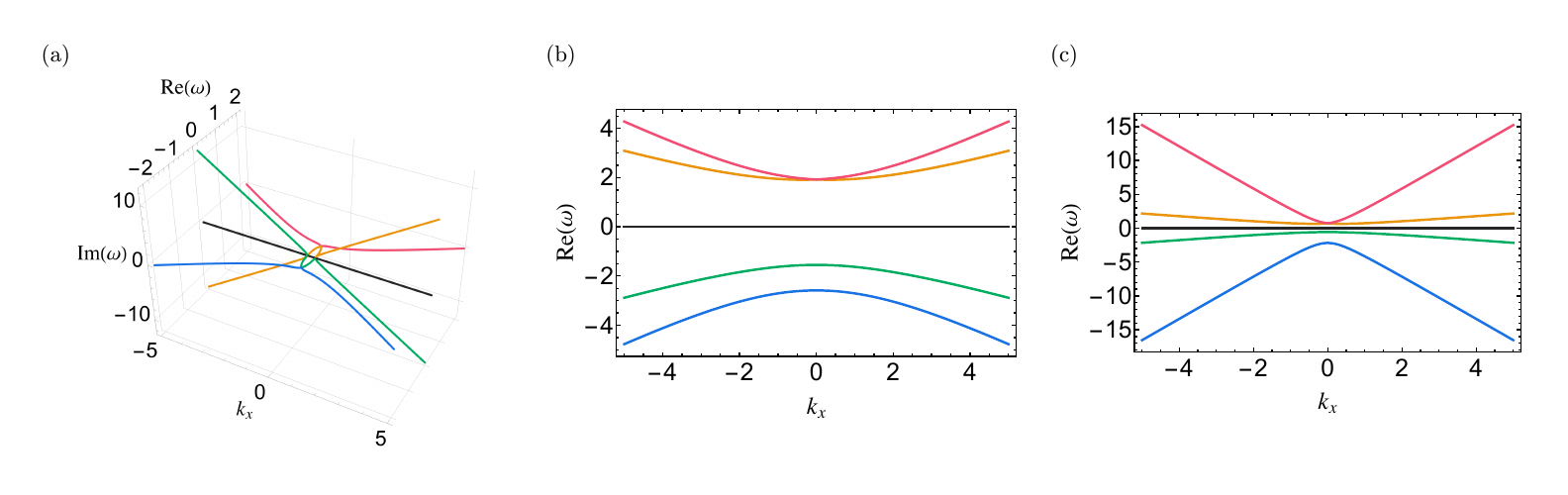}
\caption{Representative band structures along $k_x$ for the topological-transition cuts at fixed $\epsilon_{xy}=0.5$. (a): $\gamma_y=\sqrt{17/8}$ on the cut $k_y=1$, $k_z=0$, corresponding to the upper topological boundary Eq.~\eqref{sup_eq:gyro_upper_topological_curve}. (b,c): $\gamma_y=\tfrac14\sqrt{15-\sqrt{161}}$ and $\gamma_y=\tfrac14\sqrt{15+\sqrt{161}}$, with $k_y=1$ and $k_z$ chosen from Eq.~\eqref{sup_eq:gyro_lower_topological_point}, corresponding to the two solutions of Eq.~\eqref{sup_eq:gyro_lower_topological_curve} at $\epsilon_{xy}=0.5$. The band crossings at $k_x=0$ confirm the topological character of these critical lines.}
\label{fig:gyro_topo}
\end{figure}

If $k_x=0$, $k_z=0$, and $k_y\neq0$, the stationary equations give
\begin{equation}
k_x=0,
\qquad
k_y=\pm2\omega,
\qquad
k_z=0,
\label{sup_eq:gyro_upper_topological_point}
\end{equation}
and the parameter relation
\begin{equation}
\epsilon_{xy}^2=2\gamma_y^2-4,
\qquad \gamma_y^2\ge2.
\label{sup_eq:gyro_upper_topological_curve}
\end{equation}
The Hessian becomes
\begin{equation}
f_{\mathbf{k}\mathbf{k}}=\omega^2
\begin{pmatrix}
4\gamma_y^2 & 0 & 0\\
0 & 64-32\gamma_y^2 & -8\epsilon_{xy}\gamma_y\\
0 & -8\epsilon_{xy}\gamma_y & 16
\end{pmatrix},
\label{sup_eq:gyro_upper_hessian}
\end{equation}
with lower-block determinant
\begin{equation}
\det\big[(f_{\mathbf{k}\mathbf{k}})_{yz}\big]=-128(\gamma_y^2-2)(\gamma_y^2+4)\,\omega^4<0
\label{sup_eq:gyro_upper_hessian_det}
\end{equation}
for $\gamma_y^2>2$. Hence, Eq.~\eqref{sup_eq:gyro_upper_topological_curve} is the second topological transition boundary, separating $(0,\mathrm{H,II})$ from $(+2,\mathrm{H,II})$. Representative band cuts at fixed $\epsilon_{xy}=0.5$ are shown in Fig.~\ref{fig:gyro_topo}. In all three cases, the relevant bands cross at $k_x=0$, confirming the topological transitions of these critical lines.

\begin{figure}[t]
\centering
\includegraphics[width=0.8\linewidth]{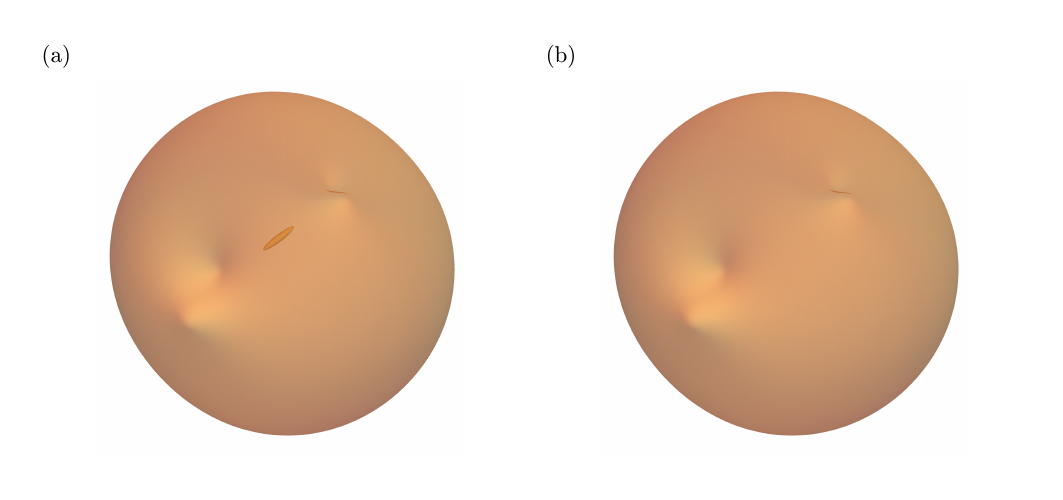}
\caption{Representative EFSs across the pinch-off Lifshitz transition at fixed $\epsilon_{xy}=0.5$. (a): $\gamma_y=1.390$, slightly below the physical branch of Eq.~\eqref{sup_eq:gyro_pinch_curve}. (b): $\gamma_y=1.392$, slightly above it. For $\epsilon_{xy}=0.5$, the theoretical pinch-off Lifshitz transition boundary lies at $\gamma_y\approx1.391$. The small pocket present on the left collapses and disappears across the boundary.}
\label{fig:gyro_pinch}
\end{figure}

If $k_x=k_y=0$ and $k_z\neq0$, the stationary equations reduce to
\begin{equation}
2\left(\frac{k_z}{\omega}\right)^3-4\left(\frac{k_z}{\omega}\right)+\epsilon_{xy}\gamma_y=0,
\label{sup_eq:gyro_pinch_axis_eq1}
\end{equation}
\begin{equation}
\left(\frac{k_z}{\omega}\right)^4-4\left(\frac{k_z}{\omega}\right)^2+2\epsilon_{xy}\gamma_y\left(\frac{k_z}{\omega}\right)-\epsilon_{xy}^2-2\gamma_y^2+4=0.
\label{sup_eq:gyro_pinch_axis_eq2}
\end{equation}
Eliminating $k_z/\omega$ yields the implicit curve
\begin{equation}
\begin{split}
&16\epsilon_{xy}^6+27\epsilon_{xy}^4\gamma_y^4-48\epsilon_{xy}^4\gamma_y^2-64\epsilon_{xy}^4
-96\epsilon_{xy}^2\gamma_y^4\\
&\hspace{1.9cm}+256\epsilon_{xy}^2\gamma_y^2+128\gamma_y^6-256\gamma_y^4=0.
\end{split}
\label{sup_eq:gyro_pinch_curve}
\end{equation}
At $(k_x,k_y,k_z)=(0,0,k_z)$ the Hessian is diagonal,
\begin{align}
\frac{\partial^2f}{\partial k_x^2}={}&\left(2\epsilon_{xy}^2+12\frac{k_z^2}{\omega^2}-24\right)\omega^2,
\label{sup_eq:gyro_pinch_hess_x}\\
\frac{\partial^2f}{\partial k_y^2}={}&\left(2\epsilon_{xy}^2-4\epsilon_{xy}\gamma_y\frac{k_z}{\omega}+12\gamma_y^2+12\frac{k_z^2}{\omega^2}-24\right)\omega^2,
\label{sup_eq:gyro_pinch_hess_y}\\
\frac{\partial^2f}{\partial k_z^2}={}&\left(48\frac{k_z^2}{\omega^2}-32\right)\omega^2.
\label{sup_eq:gyro_pinch_hess_z}
\end{align}
On the physical branch of Eq.~\eqref{sup_eq:gyro_pinch_curve}, which connects $(\epsilon_{xy},\gamma_y)=(0,\sqrt{2})$ to $(2,0)$, all three second derivatives are negative. The Hessian is therefore negative definite, so Eq.~\eqref{sup_eq:gyro_pinch_curve} is a pinch-off Lifshitz transition boundary. It separates $(0, \mathrm{E, II}^*)$ from $(0, \mathrm{E, I})$. For $\epsilon_{xy}=0.5$, the physical branch gives the critical value $\gamma_y\approx1.391$. Figure~\ref{fig:gyro_pinch} compares the EFSs at $\gamma_y=1.390$ and $1.392$: the small closed pocket present just below the boundary collapses and disappears just above it, which is the expected pinch-off evolution.

\subsection{EP transitions}
The zero-frequency EP condition is already given by Eq.~\eqref{sup_eq:gyro_zero_EP}. We now solve the finite-frequency EP ring in the same spirit as Sec.~III. Unlike the stationary families in part~b, an EP ring is determined by
\begin{equation}
 f(\omega,\mathbf{k})=0,
 \qquad
 \partial_\omega f(\omega,\mathbf{k})=0,
 \qquad \omega\neq0, \qquad \nabla_{\mathbf{k}}f\neq\mathbf{0}
\end{equation}. Because Eq.~\eqref{sup_eq:gyro_dispersion} is homogeneous of degree four in $(\omega,\mathbf{k})$, it is convenient to solve at fixed nonzero $\omega$ and introduce
\begin{equation}
 p\equiv \frac{k_x^2+k_y^2}{\omega^2},
 \qquad
 y\equiv \frac{k_y^2}{\omega^2},
 \qquad
 z\equiv \frac{k_z}{\omega}.
\label{sup_eq:gyro_EP_scaling}
\end{equation}
Then $f=0$ and $\partial_\omega f=0$ reduce to
\begin{align}
0={}&2p^2+(\epsilon_{xy}^2+6z^2-12)p-\gamma_y^2y^2+2\gamma_y(3\gamma_y-\epsilon_{xy}z)y \nonumber\\
&\hspace{1.9cm}+4\big(z^4-4z^2+2\epsilon_{xy}\gamma_y z+4-\epsilon_{xy}^2-2\gamma_y^2\big),
\label{sup_eq:gyro_EP_exact1}\\
0={}&(\epsilon_{xy}^2-12)p-\gamma_y(\epsilon_{xy}z-6\gamma_y)y \nonumber\\
&\hspace{1.9cm}-16z^2+12\epsilon_{xy}\gamma_y z-8\epsilon_{xy}^2-16\gamma_y^2+32.
\label{sup_eq:gyro_EP_exact2}
\end{align}
For \(\epsilon_{xy}^2\neq12\), Eq.~\eqref{sup_eq:gyro_EP_exact2} gives
\begin{equation}
 p=
 \frac{\gamma_y(6\gamma_y-\epsilon_{xy}z)y-16z^2+12\epsilon_{xy}\gamma_y z-8\epsilon_{xy}^2-16\gamma_y^2+32}{12-\epsilon_{xy}^2}.
\label{sup_eq:gyro_EP_p}
\end{equation}
Substituting Eq.~\eqref{sup_eq:gyro_EP_p} into Eq.~\eqref{sup_eq:gyro_EP_exact1} yields a quadratic equation for $y$,
\begin{equation}
 A(z)y^2+B(z)y+C(z)=0,
\label{sup_eq:gyro_EP_quad}
\end{equation}
with
\begin{align}
A(z)={}&-\gamma_y^2\Big[(\epsilon_{xy}^4-24\epsilon_{xy}^2-72\gamma_y^2+144)+24\epsilon_{xy}\gamma_y z-2\epsilon_{xy}^2z^2\Big],
\label{sup_eq:gyro_EP_A}\\
B(z)={}&-192\gamma_y^2(\epsilon_{xy}^2+2\gamma_y^2-4) \nonumber\\
&-\epsilon_{xy}\gamma_y(\epsilon_{xy}^4-56\epsilon_{xy}^2-352\gamma_y^2+272)z \nonumber\\
&-12\gamma_y^2(7\epsilon_{xy}^2-4)z^2+2\epsilon_{xy}\gamma_y(3\epsilon_{xy}^2-4)z^3,
\label{sup_eq:gyro_EP_B}\\
C(z)={}&4(\epsilon_{xy}^2+2\gamma_y^2-4)(\epsilon_{xy}^4+8\epsilon_{xy}^2+64\gamma_y^2+16) \nonumber\\
&-4\epsilon_{xy}\gamma_y(\epsilon_{xy}^4+72\epsilon_{xy}^2+192\gamma_y^2-240)z \nonumber\\
&+16(3\epsilon_{xy}^4+24\epsilon_{xy}^2\gamma_y^2-16\epsilon_{xy}^2-8\gamma_y^2+16)z^2 \nonumber\\
&-24\epsilon_{xy}\gamma_y(3\epsilon_{xy}^2-4)z^3+4(\epsilon_{xy}^2-4)(\epsilon_{xy}^2+4)z^4.
\label{sup_eq:gyro_EP_C}
\end{align}
Defining
\begin{equation}
\Delta(z)\equiv B(z)^2-4A(z)C(z),
\label{sup_eq:gyro_EP_discriminant}
\end{equation}
the exact finite-frequency EP ring is parameterized by
\begin{equation}
 y_{\pm}(z)=\frac{-B(z)\pm\sqrt{\Delta(z)}}{2A(z)},
 \qquad
 p_{\pm}(z)=\frac{\gamma_y(6\gamma_y-\epsilon_{xy}z)y_{\pm}(z)-16z^2+12\epsilon_{xy}\gamma_y z-8\epsilon_{xy}^2-16\gamma_y^2+32}{12-\epsilon_{xy}^2}.
\label{sup_eq:gyro_EP_general}
\end{equation}
Returning to the original wave-vector variables gives
\begin{equation}
 k_y^2=\omega^2y_{\pm}(z),
 \qquad
 k_x^2=\omega^2\big[p_{\pm}(z)-y_{\pm}(z)\big],
 \qquad
 k_z=\omega z.
\label{sup_eq:gyro_EP_kcoords}
\end{equation}
Real points on the ring require $\Delta(z)\ge0$, $y_{\pm}(z)\ge0$, and $p_{\pm}(z)\ge y_{\pm}(z)$. Equations~\eqref{sup_eq:gyro_EP_general} and \eqref{sup_eq:gyro_EP_kcoords} therefore give the general finite-frequency EP ring at fixed $\omega$. The ring is symmetric under $k_x\to-k_x$ and $k_y\to-k_y$, while the odd powers of $z$ displace it along the $k_z$ direction. At $\epsilon_{xy}^2=12$, Eq.~\eqref{sup_eq:gyro_EP_exact2} should instead be solved for $y$; this is only a parametric singularity and does not define an additional phase boundary.

\begin{figure}[t]
\centering
\includegraphics[width=0.98\linewidth]{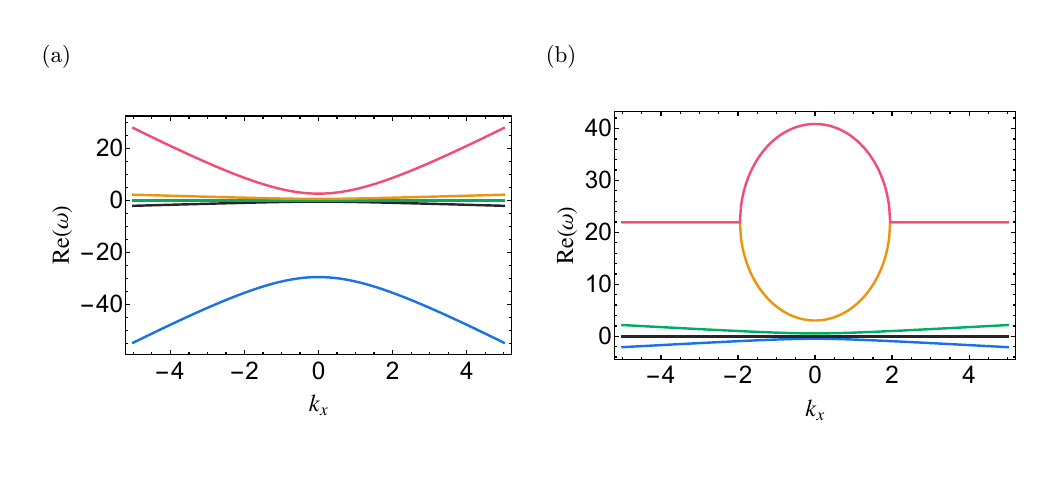}
\caption{Representative one-dimensional band cuts across the finite-frequency EP transition in the gyroelectric-chiral model at fixed $\epsilon_{xy}=0.5$, along \(\mathbf{k}=(k_x,0,1)\). (a): $\gamma_y=1.36$, slightly below the critical value $\gamma_y=\sqrt{15/8}\approx1.369$ from Eq.~\eqref{sup_eq:gyro_EP_boundary}, so no finite-frequency EP is present on the shown cut. (b): $\gamma_y=1.375$, slightly above the boundary; the cut intersects the finite-frequency EP ring and shows two finite-frequency EPs at $k_x\approx\pm1.95$ and $\mathrm{Re}\,\omega\approx22.0$.}
\label{fig:gyro_ep}
\end{figure}

The transition boundary is obtained when the ring contracts to the origin, i.e., $p=y=z=0$. Equations~\eqref{sup_eq:gyro_EP_exact1} and \eqref{sup_eq:gyro_EP_exact2} then give
\begin{equation}
\epsilon_{xy}^2+2\gamma_y^2=4,
\label{sup_eq:gyro_EP_boundary}
\end{equation}
which is exactly the energy-metric singularity in Eq.~\eqref{sup_eq:gyro_metric_curve}. At $\mathbf{k}=0$, the $K$ matrix is
\begin{equation}
K(\omega,\mathbf{0})=\omega^2
\begin{pmatrix}
2 & i\epsilon_{xy} & 0\\
-i\epsilon_{xy} & 2-\gamma_y^2 & 0\\
0 & 0 & 4
\end{pmatrix}.
\label{sup_eq:gyro_K_origin}
\end{equation}
We can easily get
\begin{equation}
\mathrm{rank}\,K=2,
\qquad
\mathrm{GM}=1.
\label{sup_eq:gyro_rank_origin}
\end{equation}
Moreover,
\begin{equation}
\nabla_{\mathbf{k}}f(\omega,\mathbf{0})=(0,0,8\epsilon_{xy}\gamma_y\omega^3)\neq\mathbf{0},
\label{sup_eq:gyro_grad_origin}
\end{equation}
Hence Eq.~\eqref{sup_eq:gyro_EP_boundary} is the finite-frequency EP-transition boundary. At fixed $\epsilon_{xy}=0.5$, Eq.~\eqref{sup_eq:gyro_EP_boundary} gives the critical value $\gamma_y=\sqrt{15/8}\approx1.369$. Representative one-dimensional band cuts on the two sides of this boundary are shown in Fig.~\ref{fig:gyro_ep}.

\subsection{Phase diagram summary}
Collecting the results above reconstructs the phase diagram in Fig.~2 of the main text. Each phase is identified by the composite label $(\mathcal{C}, \mathrm{Geometry, Type})$, where $\mathcal{C}$ represents the Chern number of the highest band ($-2$, $0$, or $+2$), and the second entry denotes the EFS geometry ($\mathrm{E}/\mathrm{H}$ for elliptical (closed)/hyperbolic (open), and $\mathrm{I}/\mathrm{II}$ for one-sheet/two-sheet). Additionally, $(\mathrm{E, II}^*)$ designates a special $(\mathrm{E, II})$ phase where the origin is geometrically excluded from the EFS interior. The six sectors are:
\begin{itemize}
\item \textbf{$(-2, \mathrm{E, II})$}: $0\le\gamma_y^2<2$, below the lower topological boundary in Eq.~\eqref{sup_eq:gyro_lower_topological_curve}.
\item \textbf{$(0, \mathrm{E, II})$}: $0\le\gamma_y^2<2$, above Eq.~\eqref{sup_eq:gyro_lower_topological_curve} but inside the ellipse Eq.~\eqref{sup_eq:gyro_metric_curve}.
\item \textbf{$(0, \mathrm{E, II}^*)$}: $0\le\gamma_y^2<2$, outside Eq.~\eqref{sup_eq:gyro_metric_curve} but before the pinch-off curve Eq.~\eqref{sup_eq:gyro_pinch_curve}; this is the closed-sheet sector with the exact finite-frequency EP ring in Eqs.~\eqref{sup_eq:gyro_EP_general} and \eqref{sup_eq:gyro_EP_kcoords}, born on Eq.~\eqref{sup_eq:gyro_EP_boundary}.
\item \textbf{$(0, \mathrm{E, I})$}: the remaining closed-sheet, one-sheet, Chern-$0$ sector beyond the pinch-off curve Eq.~\eqref{sup_eq:gyro_pinch_curve}.
\item \textbf{$(0, \mathrm{H, II})$}: $\gamma_y^2>2$, and below the upper topological boundary in Eq.~\eqref{sup_eq:gyro_upper_topological_curve}; this is the open-sheet Chern-$0$ sector.
\item \textbf{$(+2, \mathrm{H, II})$}: above Eq.~\eqref{sup_eq:gyro_upper_topological_curve}; this is the open-sheet Chern-$+2$ sector.
\end{itemize}

\bibliography{ref}